\documentclass[aps,prb,10pt,twocolumn,floatfix,showpacs,superscriptaddress,longbibliography]{revtex4-2}
\usepackage[dvipsnames]{xcolor}
\usepackage{amsmath}
\usepackage{graphicx}
\usepackage[colorlinks=true,citecolor=blue,linkcolor=blue,urlcolor=blue]{hyperref}
\usepackage{placeins}
\usepackage[T1]{fontenc}
\usepackage{dcolumn}% Align table columns on decimal point
\usepackage{bm}% bold math

\newcommand{\avg}[1]{\big< #1 \big>} % for average
\newcommand{\lr}[1]{\left(#1\right)}

\newcommand{\cC}[1]{c^{\dagger}_{#1}}
\newcommand{\cA}[1]{c_{#1}}

\begin{document}
\title{Relaxation dynamics of a quantum spin coupled to a topological edge state}
\author{Qiyu Liu}
 \affiliation{Technische Universität Braunschweig, Institut für Mathematische Physik}
\author{Christoph Karrasch}
 \affiliation{Technische Universität Braunschweig, Institut für Mathematische Physik} 
 \author{Dante Marvin Kennes}
\affiliation{Institut f\"ur Theorie der Statistischen Physik, RWTH Aachen University and JARA-Fundamentals of Future Information Technology, 52056 Aachen, Germany}
\affiliation{Max Planck Institute for the Structure and Dynamics of Matter, Center for Free-Electron Laser Science, 22761 Hamburg, Germany}
 \author{Roman Rausch}
 \affiliation{Technische Universität Braunschweig, Institut für Mathematische Physik}
\date{\today}% It is always \today, today,

\begin{abstract}
A classical impurity spin coupled to the spinful Su-Schrieffer-Heeger (SSH) chain is known to exhibit complex switching dynamics with incomplete spin relaxation. Here, we study the corrections that result from a full quantum treatment of the impurity spin. We find that in the topologically trivial case, the quantum spin behaves similarly to the classical one due to the absence of the Kondo effect for the trivial insulator.
In the topological case, the quantum spin is significantly less likely to relax: It can be stuck at a pre-relaxation plateau with a sizable deviation from the expected relaxed value, and there is a large parameter regime where it does not relax at all but features an anomalously large Larmor frequency. Furthermore, we find an additional quantum effect where the pre-relaxation plateau can be hyperpolarized, i.e., the spin is stuck at a polarization value larger than the ground-state expectation value. This is possible due to the (incomplete) Kondo screening of the quantum spin, which is absent in the classical case. Our results are obtained via the ground state density matrix renormalization group (DMRG) algorithm and the time-dependent variational principle (TDVP), where the charge-SU(2) symmetry of the problem was exploited. Furthermore, we introduce and benchmark a method to predict the dynamics from the given numerical data based on the sparse identification of nonlinear dynamics (SINDy). This allows us to prolong the simulation timescale by a factor of 2.5, up to a maximal time of $10^3$ inverse hoppings.
\end{abstract}

\maketitle
\section{Introduction}

A quantum spin is the basic building block of systems that can be turned into quantum devices. The spin can be manipulated by external fields, but interactions with an environment can lead to eventual decoherence and relaxation.
The understanding of this dynamics is an essential prerequisite for the development of spintronics-based devices, where the spin is used as a carrier of information, but is also relevant for the rapid recent developments in quantum information theory.
From a more fundamental point of view, it poses a case study of the physics of open quantum systems and lifts the analysis of quantum impurity models, which has an extensive history~\cite{Kondo1969,Wilson1975}, into the nonequilibrium domain~\cite{Kohn2021,Thoenniss2023}.

As with all time-dependent problems, the main technical challenge in understanding the spin dynamics is to access long propagation times. While the direct exchange interaction is on the scale of femtoseconds, nontrivial dynamics can occur over a wide range of time scales that can be several orders of magnitude larger. The study of impurity models in equilibrium has stimulated the development of novel techniques like the numerical renormalization group (NRG)~\cite{Wilson1975}, but accessing  the non-equlibrium dynamics of a quantum spin attached to an environment remains a difficult problem~\cite{Costi1997,Nordlander1999,Lechtenberg2014}.

In addition to tailoring external fields, controlling the environment provides another route to manipulate the dynamics of the spin. In the following, we focus on the case of fermionic baths (substrates), where a particularly interesting case arises when the spin is coupled to the robust surface state of a topological insulator. One expects that the resulting spin dynamics should differ fundamentally compared to both metallic and topologically trivial substrates. The authors of Ref.~\cite{Bouaziz2019} investigated magnetic impurities embedded in Bi$_2$Te$_3$ and Bi$_2$Se$_3$ using time-dependent density-functional theory and found a wide range of excitation lifetimes, reaching up to microseconds for Mn impurities. Using Mn adatoms on Bi$_2$Se$_3$ has also been proposed as a building block of a magnetic sensor device~\cite{Narayan2015}. The effective interaction of several magnetic impurities mediated by topological edge states has been investigated using first-principle methods~\cite{Chotorlishvili2014}. Analogous protection and engineered response have recently been demonstrated in photonic topological platforms, with robust polarization conversion in anisotropic metamaterial bilayers~\cite{Sarsen2019} and ultralow-frequency topological scattering resonances in magnetised core-shell plasmonic cylinders~\cite{Gangaraj2020}.

While the two- and three-dimensional cases are most interesting from an applicational point of view, the one-dimensional case is useful in elucidating the fundamental physics with few or even no additional approximations. One way to model a topological substrate is through the paradigmatic Su-Schrieffer-Heeger (SSH) model~\cite{Su1980,Heeger1988}. A recent study investigated this setup for a classical instead of a quantum spin, which results in an effective one-particle problem and makes long-time dynamics tractable~\cite{Elbracht2021}. By virtue of absorbing Lindblad boundaries, the authors were able to reach timescales which exceed the intrinsic electronic one (set by the inverse hopping amplitude) by a factor of $10^{5}$. It was observed that an in-gap edge state facilitates the spin relaxation in a wider parameter range of the dynamical phase diagram as compared to the topologically trivial case. However, the relaxation process generally remains incomplete and the spin is stuck at a few percent below its relaxed value for very long times.

In this work, we build on these results and investigate what happens when the spin is faithfully treated as a quantum object with $S=1/2$. This is an interesting question because a quantum spin has the fundamental property that the substrate electrons can truly screen it via the Kondo effect, up to a total singlet in the metallic case, instead of just aligning themselves antiferromagnetically to it.
In the metallic and topologically nontrivial cases, the competition of the Kondo effect and a local external field leads to deviation from the saturated magnetization value already in equilibrium. In this paper, we switch the direction of the magnetic field at time $t=0$ and study the ensuing non-equilibrium dynamics. It turns out that the quantum spin is generally less likely to relax than a classical one and that it features hyperpolarization, i.e., its polarization can exceed the equilibrium value and remain stuck in this pre-relaxed state.

We compute the ground state of the system at hand using the density-matrix renormalization group (DMRG) method and simulate the ensuing dynamics via the time-dependent variational principle (TDVP)~\cite{Haegeman2016}. By exploiting the charge-SU(2) symmetry of the problem, this allows us to reach timescales of $\sim400$ inverse hoppings. We extend this timescale further by employing the sparse identification of nonlinear dynamics (SINDy)~\cite{Brunton2016,Quade2018,Kaheman2020,Abdullah2023}, which attempts to learn the subsequent dynamics from the data at earlier times. This allows us to gain an additional factor of $2.5$, bringing the maximally achieved propagation time to $10^{3}$ inverse hoppings.

%%%%%%%%%%%%%%%%%%%
\section{\label{section:model}Model}
%%%%%%%%%%%%%%%%%%%

\begin{figure}
\includegraphics[width=0.8\columnwidth]{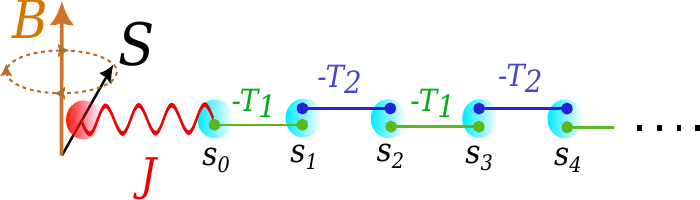}
\caption{Sketch of the system studied in this work. An impurity spin $\boldsymbol{S}$ is coupled via an exchange interaction $J$ to the edge of a SSH chain whose hoppings are dimerized as $T_1=T-\delta T$ and $T_2=T+\delta T$. The spin is switched by an external field $\boldsymbol{B}(t)$.
}
\label{fig:figureofmodel}
\end{figure}

We investigate the following Hamiltonian (see Fig.~\ref{fig:figureofmodel}):
\begin{equation}
\begin{split}
    H &=  -\sum_{j=0}^{L-2}\sum_\sigma\Big[ \big(T - (-1)^{j}\delta T \big) c^{\dagger}_{j,\sigma} c_{j+1,\sigma} +\text{h.c.} \Big]\\
    &+J\boldsymbol{s}_{0}\cdot\boldsymbol{S} - \boldsymbol{B}(t)\cdot\boldsymbol{S},
\label{eq:generic_hamiltonian}
\end{split}
\end{equation}
where $c^{\dagger}_{j,\sigma}$ creates an electron with spin projection $\sigma=\uparrow,\downarrow$ at site $j$, and $\boldsymbol{s}_{j} = \frac{1}{2} \sum_{\sigma\sigma'} c^{\dagger}_{j,\sigma} \boldsymbol{\tau}_{\sigma\sigma'} c_{j,\sigma'}$ is its spin operator constructed using the vector of Pauli matrices $\boldsymbol{\tau}$. $T$ is the nearest-neighbor hopping amplitude, which we set to $T=1$ from now on. The uncoupled system ($J=0$) is in a topologically nontrivial phase for $\delta T>0$. Note that the bulk gap is given by $\Delta=4\big|\delta T\big|$.

The impurity spin is described by the vector of spin operators $\boldsymbol{S}=(S^x,S^y,S^z)$. It is coupled to the first site of the electronic substrate via the exchange interaction $J>0$  and is manipulated by the external field $\boldsymbol{B}(t)=(B^x(t),B^y(t),B^z(t))$. We set $J=1$.

In this work, we study switching dynamics, i.e., we prepare the system in the ground state with
\begin{equation}
\boldsymbol{B}(t=0)=(B,0,0)
\end{equation}
and simulate the ensuing time evolution with $\boldsymbol{B}(t>0)=(0,0,B)$. Since the exchange interaction is isotropic, one expects that the system relaxes towards a state where the $x$ and $z$ components are exchanged, $\avg{S^{x,z}}_{\text{relax}}=\avg{S^{z,x}(t=0)}$.

We use finite chains of length $L=401$. The switching puts additional energy into the system which is transported away in the substrate with some finite speed. This wavefront is reflected at the opposite boundary and moves back to the impurity spin. The simulation should be stopped before the spin is influenced by this finite-size effect. This sets a maximal propagation time, which can be estimated from the dispersion relation, see App.~\ref{app:proptime}.

%%%%%%%%%%%%%%%%%%%%%%%
\section{\label{section:method}Methods} 
%%%%%%%%%%%%%%%%%%%%%%%

\subsection{DMRG, TDVP} We employ the DMRG algorithm to first compute the ground state of Eq.~\eqref{eq:generic_hamiltonian}. The DMRG is a variational method within the class of variational matrix-product states (MPS) and is especially efficient in 1D due to the area law of the entanglement entropy. The MPS ansatz is controlled by the bond dimension $\chi$, which is related to the number of free parameters in the ansatz.

By exploiting symmetries, the MPS tensors get a block structure, so that a larger effective $\chi$ can be accessed. For our problem, the ground state features a spin-U(1) symmetry, but it is broken during the time evolution by the switching of the magnetic field.
Another symmetry is particle number conservation, which is in fact part of an even larger charge-SU(2) symmetry~\cite{Zhang1990,Essler2005}. The prerequisite for this is the bipartiteness of the lattice, which allows to define the phase factors $(-1)^j = \pm 1$ for the respective sublattices. The symmetry is generated by the pseudospin operators
\begin{eqnarray}
T^+_j &=& \lr{-1}^j \cA{j,\downarrow} \cA{j,\uparrow}, \label{eq:Tdef}\\
T^-_j &=& \lr{-1}^j \cC{j,\uparrow} \cC{j,\downarrow}, \nonumber\\
T^z_j &=& \frac{1}{2}\lr{ 1- \cC{j,\uparrow}\cA{j,\uparrow}-\cC{j,\downarrow}\cA{j,\downarrow}} ,\nonumber
\end{eqnarray}
which fulfill the SU(2) relations $[T^+_j,T^-_j]=2T^z_j$ and $[T^z_j,T^{\pm}_j]=\pm T^{\pm}_j$. The global operators $T^x_{\text{tot}}=\sum_j\lr{T^+_j+T^-_j}/2$, $T^y_{\text{tot}}=\sum_j\lr{T^+_j-T^-_j}/(2i)$, $T^z_{\text{tot}}=\sum_jT^z_j$ all commute with the Hamiltonian. Using ${\bf T}_i = (T^x_i,T^y_i,T^z_i)$, the total pseudospin is defined via
\begin{equation}
\big\langle{\bf T}_\text{tot}^2\big\rangle = \sum_{ij} \big\langle{\bf T}_i\cdot{\bf T}_j\big\rangle = T_\text{tot}\lr{T_\text{tot}+1}.
\end{equation}
We always work at half filling $\langle T^z_\text{tot}\rangle=0$, and the ground state is always in the sector $T_\text{tot}=0$, which we have verified exemplarily. By exploiting the full charge-SU(2) symmetry \cite{McCulloch2002}, we can effectively increase the bond dimension by a factor of $2.5$ compared to the case of charge-U(1) conservation alone (which corresponds to just the pseudospin projection $T^z_\text{tot}$).

We use a charge-SU(2) effective bond dimension $\chi_{\text{SU(2)}}=150$ to compute the ground state using a mixture of the two-site DMRG algorithm and the one-site algorithm with perturbations~\cite{Hubig2015}. The resulting energy variance per site $\text{var}(E)/L=\big|\avg{H^2} -\avg{H}^2\big|/L$ is of the order of $\text{var}(E)/L\sim10^{-5}$, indicating a well-converged ground state. For the time propagation, we employ an adaptive variant of the time-dependent variational principle~\cite{Haegeman2016}, where we dynamically choose whether the two-site or the one-site algorithm should be performed on a given site. We use the two-site algorithm close to the impurity spin and around the moving wavefront, which we detect by the relative growth of the entanglement entropy. The two-site algorithm is able to dynamically increase the bond dimension in order to faithfully describe the evolving state. We use a local truncated weight threshold of $10^{-7}$ and a timestep of $dt=0.1$. A benchmark with different thresholds is shown in App.~\ref{app:accuracy}.
We observe that the dynamics of the Kondo spin does not lead to a substantial growth of the entanglement, and the bond dimension does not need to increase beyond $\chi_\text{SU(2)}=150$ to faithfully capture the time evolution. In essence, one only needs to capture the outgoing wavepacket and the rotation of the impurity spin. This motivates us to investigate whether the dynamics can be learned and captured by a cheaper method, which we describe in the following.

\begin{figure}
\includegraphics[width=1.0\columnwidth]{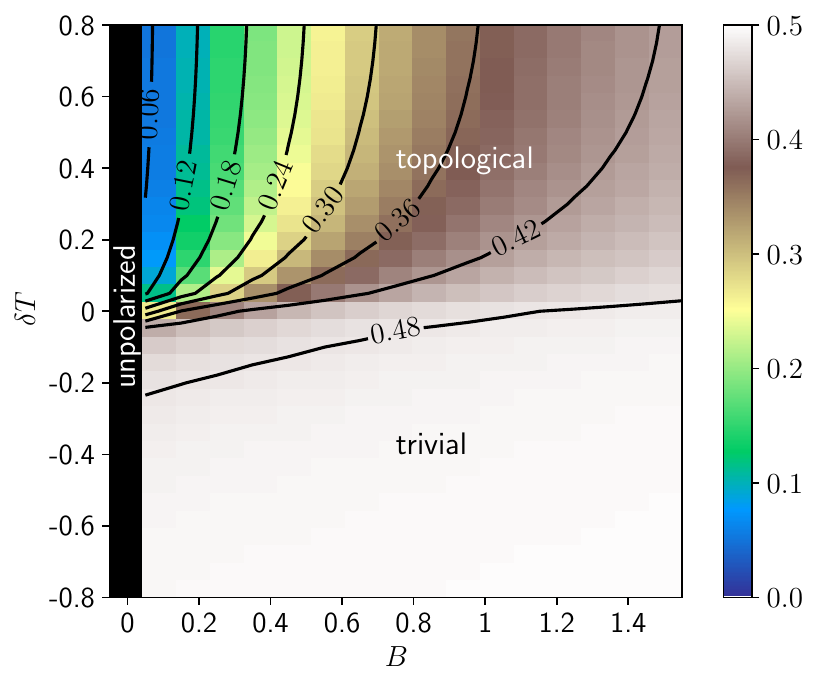}
\caption{\label{fig:groundstatePhasediagram}
Ground-state expectation value of the impurity spin $\left<S^x\right>$ for $L=401$, $T=J=1$, and $\boldsymbol{B}=(B,0,0)$ computed using the DMRG at half filling.
}
\end{figure}

\subsection{SINDy}
In order to access larger timescales after the TDVP simulation has terminated, we employ the SINDy algorithm, which is a data-driven scheme to learn and predict the dynamics. It can be combined with any time evolution technique.

A physical motivation for the SINDy approach can be formulated as follows: A known classical differential equation with phenomenological parameters to describe spin dynamics is the Landau-Lifshitz-Gilbert (LLG) equation~\cite{Breuer2002,Fransson2008,Bhattacharjee2012}. It is related to the microscopic Hamiltonian via a Redfield Master Equation requiring the weak-coupling, Markov, and classical-spin approximations, leading to the following form~\cite{Sayad2015,Vedmedenko2018}:
\begin{equation}
\dot{\mathbf{S}}(t) = \mathbf{S}(t)\times\mathbf{B} - \alpha \mathbf{S}(t)\times\dot{\mathbf{S}}(t).
\label{eq:LLG}
\end{equation}
For the classical-spin case with fully included electronic degrees of freedom, its validity has been checked against the full time propagation~\cite{Elbracht2024}.

In this work, we do not make the above assumptions but rather compute the exact dynamics using the TDVP up to a maximal time. We then assume that one can find an effective classical equation for the expectation value $\avg{\mathbf{S}(t)}$, which takes a more complex form than the LLG equation. This equation can be found by the SINDy method~\cite{Brunton2016,Quade2018,Kaheman2020,Abdullah2023}, which determines a nonlinear function $f$ that governs the time evolution of a set of $m$ observables $\boldsymbol{A}_i(t)$ via
\begin{equation}
\frac{d}{dt}\boldsymbol{A}_i(t) =  f\Big(\big\{\boldsymbol{A}_j(t)\big\} \Big).
\label{eq:definition of SINDy}
\end{equation}
In practice, one first performs a set of measurements of the system at times $t_1, t_2, \dots, t_n$ and then aggregates these data into an $n\times m$ matrix $\underline{X}$:
\begin{equation}
\underline{X} = 
\left[ \begin{array}{cccc}
\boldsymbol{A}_1(t_1) & \boldsymbol{A}_2(t_1) & \dots & \boldsymbol{A}_m(t_1) \\
\boldsymbol{A}_1(t_2) & \boldsymbol{A}_2(t_2)  &\dots & \boldsymbol{A}_m(t_2) \\
\vdots & \vdots &  & \vdots \\
\boldsymbol{A}_1(t_n) & \boldsymbol{A}_2(t_n)  &\dots & \boldsymbol{A}_m(t_n) \\
\end{array}\right]
\label{eq:data matrix}.
\end{equation}
Given $\underline{X}$, time derivatives can be approximated numerically using spline interpolation and reshaped into a matrix $\underline{\dot{X}}$. After this, a set of basis functions to form a library matrix $\underline{\Theta}(\underline{X})$ is chosen, such as polynomials up to order $k$:
\begin{equation}
    \underline{\Theta}(X) = 
  \begin{bmatrix}
    \underline{P}_0(\underline{X}) &  \underline{P_1}(\underline{X}) &  \underline{P_2}(\underline{X})&\dots&\underline{P_k}(\underline{X})
  \end{bmatrix}
\label{eq:polynomial basis}.
\end{equation}
For example, a term with $k=2$ takes the form
\begin{equation}
    \underline{P_2}(\underline{X}) = 
  \begin{bmatrix}
    \mid &\mid & &\mid &\mid &  &\mid \\
    \boldsymbol{A}_1^2 & \boldsymbol{A}_1\boldsymbol{A}_2 &  \dots&\boldsymbol{A}_1\boldsymbol{A}_m&\boldsymbol{A}_2^2&\dots&\boldsymbol{A}_m^2\\
        \mid &\mid & &\mid &\mid &  &\mid 
  \end{bmatrix}
\label{eq:second order polynomial basis}.
\end{equation}
Given $\underline{\dot{X}}$ and $\underline{\Theta}(X)$, one obtains a set of coefficients for the right-hand side of Eq.~\eqref{eq:definition of SINDy} with a matrix form $\underline{\Xi}$ by solving the following approximation problem:
\begin{equation}
    \dot{\underline{X}} \approx \underline{\Theta}(X)\,\underline{\Xi}.
\label{eq:SINDy approximation problmes}
\end{equation}
Crucially, the set of coefficients is sparse, involving less functions than the full library.

The simplest way to apply the SINDy algorithm to our system is to set $\boldsymbol{A}(t)=\avg{\boldsymbol{S}(t)}$ and to learn an effective differential equation of the impurity spin. We find that this works rather well for the simple case of a metallic substrate, but fails for more complex dynamics. In the latter case, one needs to take more observables into account, which we discuss below.

\begin{figure}
\includegraphics[width=0.95\columnwidth]{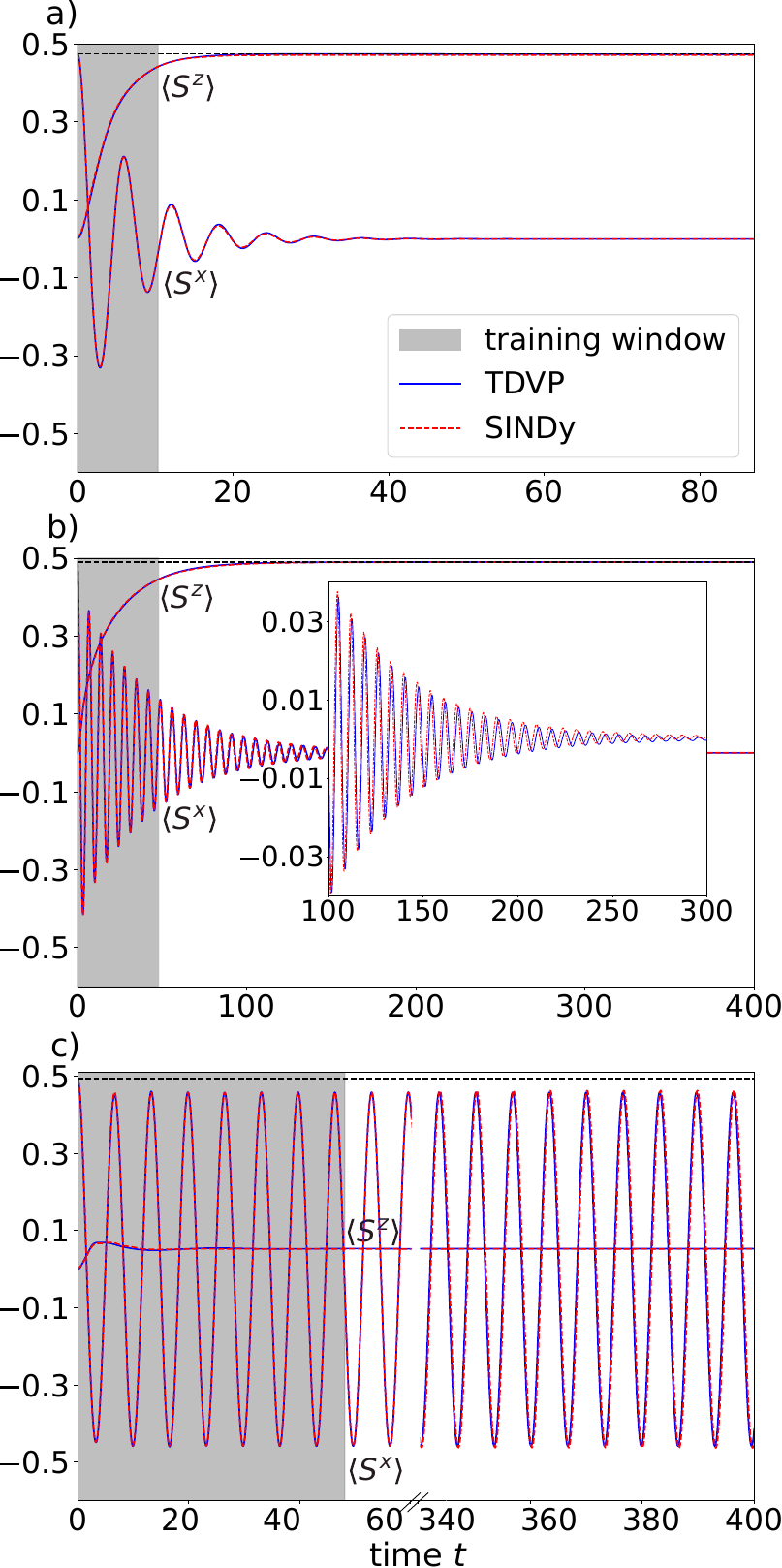}
\caption{\label{fig:metallic and trivial cases}
Relaxation dynamics of the impurity spin for $B=1$: (a) metallic case ($\delta T =0$), (b) topologically trivial case with relaxation ($\delta T =-0.2\Rightarrow\Delta=0.8<\omega_L\approx B$), (c) topologically trivial case without relaxation ($\delta T =-0.3\Rightarrow\Delta=1.2>\omega_L\approx B$).
The gray shading indicates the learning time window for the SINDy method. Note that in b) and c) larger time windows are considered than in a).
}
\end{figure}

\begin{figure*}
    \centering
    \includegraphics[width=2.0\columnwidth]{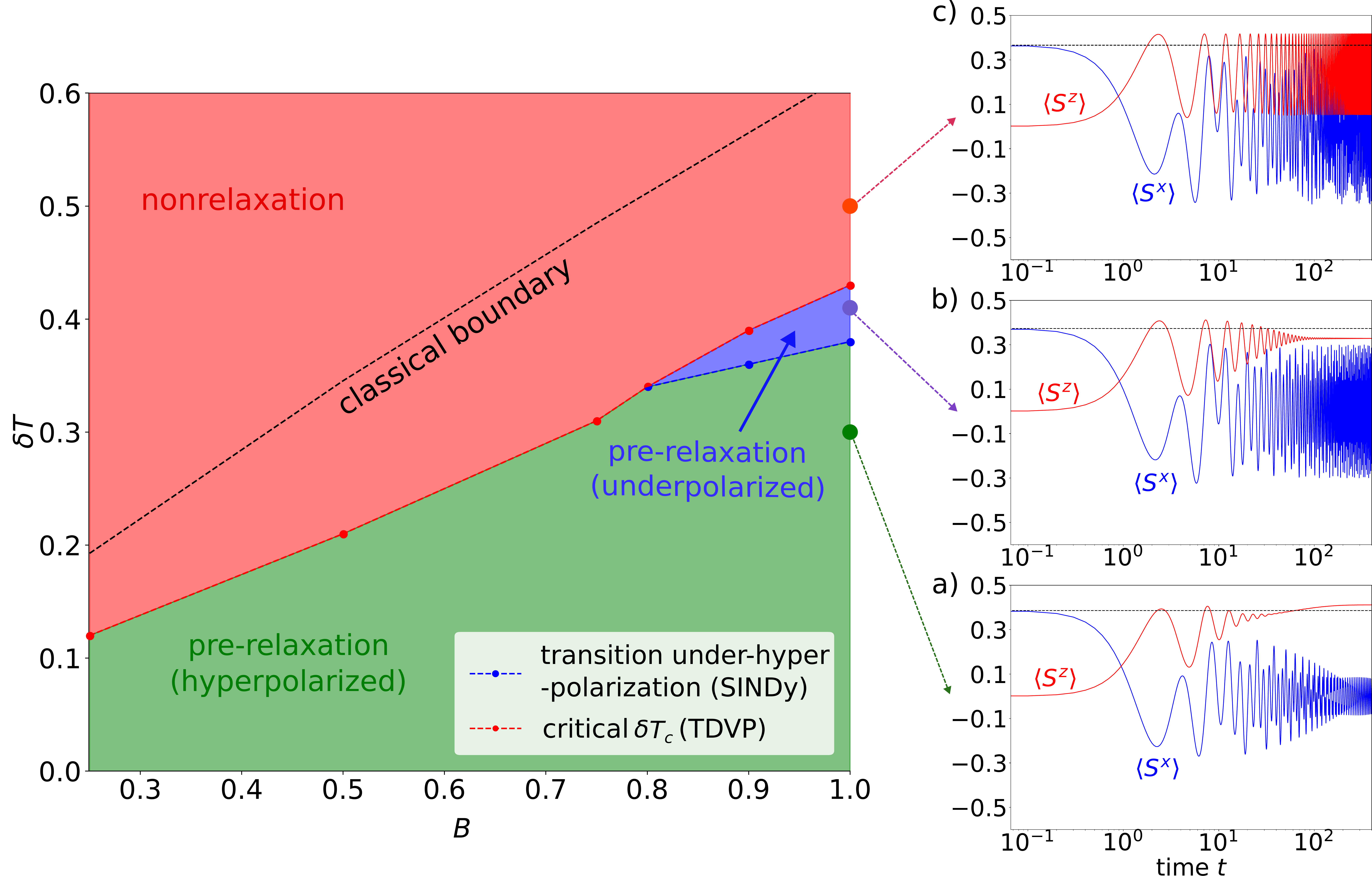}
\caption{\label{fig:regimes}
Relaxation dynamics of the impurity spin in the topological case. Left: Schematic dynamical phase diagram as a function of $B$ and $\delta T$. We find three distinct regimes; exemplary TDVP data for the time evolution at $B=1$ is shown on the right. (a) pre-relaxation and hyperpolarization (green, $\delta T=0.3$), (b) pre-relaxation and underpolarization (blue, $\delta T=0.41$), and (c) nonrelaxation (red, $\delta T=0.5$). The dotted horizontal lines indicate the expected relaxation value $\avg{S^z}_{\text{relax}}=\avg{S^x(t=0)}$. For a classical spin, nonrelaxation occurs above the black dashed line.
}
\end{figure*}

%%%%%%%%%
\section{Results}
%%%%%%%%%

\subsection{Ground state}

Figure~\ref{fig:groundstatePhasediagram} shows the polarization of the impurity spin $\avg{S^x(t=0)}$ in the ground state as a function of the magnetic field $B$. The Kondo effect is not active for the trivial insulator ($\delta T<0$), where the electrons are inert and the impurity spin is immediately fully polarized irrespective of the field strength $B$, resembling a classical spin. For the topological insulator ($\delta T>0$), the edge state is able to screen the spin and a compromise state sets in, whereby the polarization is rather small even for weak fields.

%%%%%%%%%%%%%%%%%
\subsection{Relaxation dynamics: metal and trivial insulator}
%%%%%%%%%%%%%%%%%

We now turn to the dynamics; we re-iterate that the direction of the magnetic field is switched at time $t=0$. We first look at the simple cases of a metal and a trivial insulator. There are three dynamical regimes which are shown in Fig.~\ref{fig:metallic and trivial cases} for $B=1$. The metallic case  ($\delta T=0$) has been investigated before~\cite{Sayad2016} and is reproduced here for completeness. One finds a precessional motion with a Larmor frequency of $\omega_L \approx B$, which is damped with a relaxation time of around $\tau\sim 20$ for the $x$ component.
For the trivial insulator, the Larmor frequency has to be compared with the gap size: For $\Delta=4\big|\delta T\big|<\omega_L$, electrons can be excited across the gap and the behavior qualitatively resembles the metal, albeit with a longer relaxation time of around $\tau\sim 100$.
For $\Delta=4\big|\delta T\big|>\omega_L$, the gap cannot be surpassed and one finds precessional motion without any relaxation.

We now apply and benchmark the SINDy method. The dynamics in the metallic case can be learned readily from only $\boldsymbol{A}(t)=\avg{\boldsymbol{S}(t)}$. We vary the learning window (shaded gray area in Fig.~\ref{fig:metallic and trivial cases}) until the rest of the dynamics is satisfactorily reproduced; $t_{\text{learn}}\approx 13$, i.e., about half of the relaxation time, is enough to make an accurate prediction.
For the insulating case, we find that the dynamics cannot be learned with $\avg{\boldsymbol{S}(t)}$ alone, one also needs to include the first substrate spin $\avg{\boldsymbol{s}_0(t)}$. We find $t_{\text{learn}}\approx 50$ to be sufficient, which is still about half of the relaxation time.
Our interpretation is that in the insulating case, a strongly coupled two-spin system is formed at the edge, whereas in the metallic case, the substrate merely acts as an absorbing bath that can be effectively integrated out.

%%%%%%%%%%%%%%%%%%%
\subsection{Relaxation process: topological insulator}
%%%%%%%%%%%%%%%%%%%

For the topological insulator, we can identify three distinct dynamic regimes. They are displayed in Fig.~\ref{fig:regimes} as a full phase diagram, and exemplary TDVP data is shown at $B=1$:
(a) For a small value of $\delta T>0$, the $z$ component of the impurity spin surpasses the expected relaxation value $\avg{S^z}_{\text{relax}}=\avg{S^x(t=0)}$ and stays in this hyperpolarized state without any clear signature of relaxation in the given observation time.
(b) For intermediate values of $\delta T$, the spin reaches a pre-relaxation plateau below the expected relaxed value. This is similar to the case of a classical spin, but with a larger deviation from the relaxed value and an additional superimposed high-frequency oscillation.
(c) For even larger $\delta T$, the gap is too large to overcome and strong oscillations without relaxation are found in all components. In the following, we discuss and analyze these results in more detail. We also illustrate how to quantify the above values in order to establish the phase diagram in Fig.~\ref{fig:regimes}.

In order to quantitatively discern the nonrelaxation regime (c) from the pre-relaxation regimes (a) and (b), we compare the amplitude of the oscillations of $\langle S^z(t)\rangle$ at late and early times. We define the early time oscillation as the full period immediately following the first maximum of $\langle S^z(t)\rangle$, and the late time oscillation as the last complete period before the longest simulation time achievable via TDVP. The phase boundary $\delta T_c$ in Fig.~\ref{fig:regimes} (red dashed line) shows the parameters for which their ratio is 1/20. Note that the nonrelaxation regime is larger than in the case of a classical spin (see the black dotted line in Fig.~\ref{fig:regimes}).

\begin{figure}
\includegraphics[width=1\columnwidth]{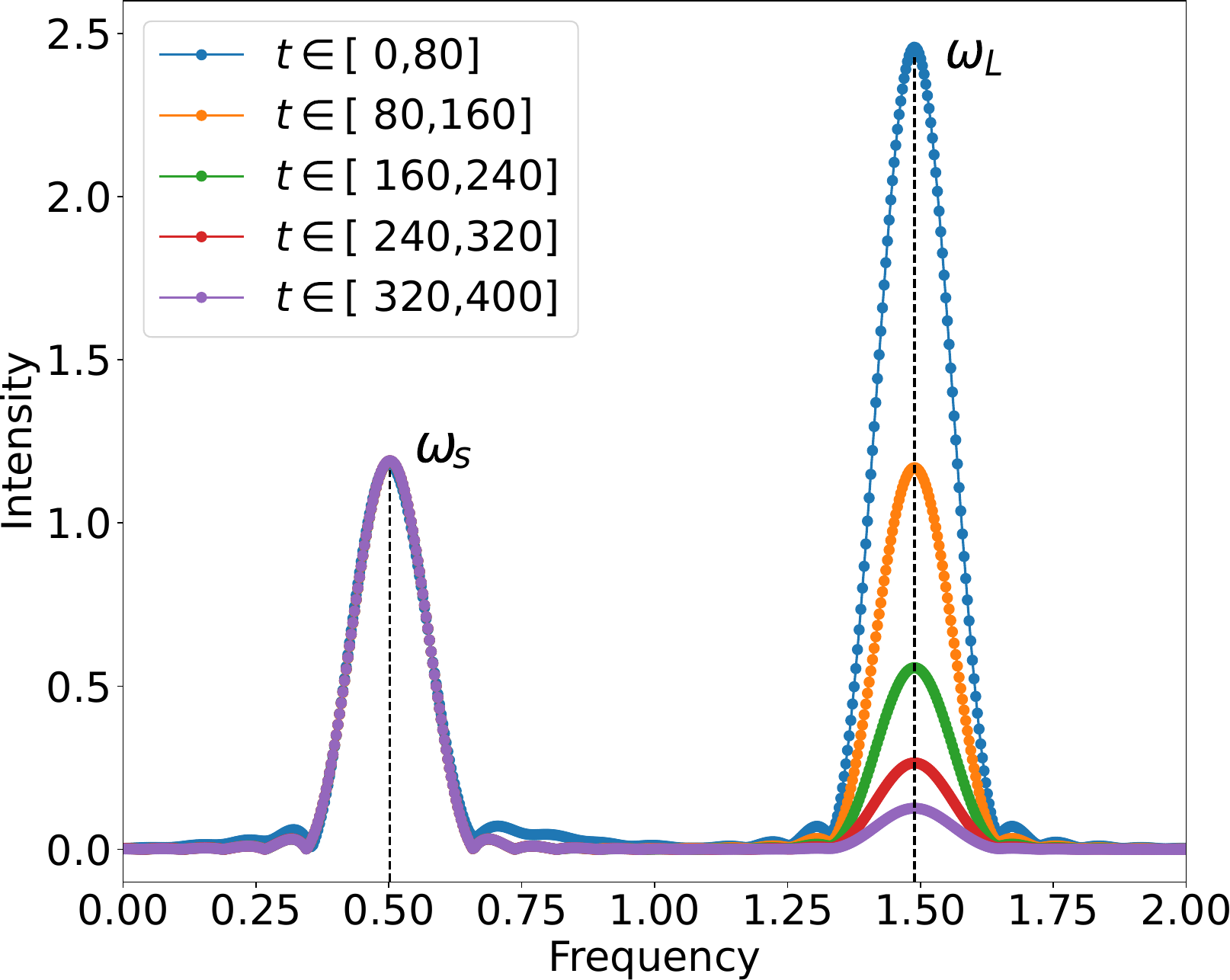}
\caption{Fourier analysis of the TDVP data for $\avg{S^x(t)}$ in the topologically non-trivial case $\delta T = 0.3$ at $B=1$. The higher and lower frequencies are labelled $\omega_L$ and $\omega_S$, respectively. By varying the time window used for the analysis, one can estimate the damping of the corresponding oscillations.}
\label{fig:Freq_Simp_delta}
\end{figure}

Next, we carry out a Fourier analysis of the TDVP data for $\langle S^x(t)\rangle$ at $B=1$ and not too large $\delta T$ and find two main frequencies, see Fig.~\ref{fig:Freq_Simp_delta}. From the equation of motion for the impurity spin,
\begin{equation}
    \langle \dot{\boldsymbol{S}} \rangle = -i \langle [\boldsymbol{S}, H]  \rangle = -\boldsymbol{B} \times \langle \boldsymbol{S} \rangle + J\langle \boldsymbol{s}_0 \times \boldsymbol{S} \rangle,
\label{eq: equation of motion of impurity spin}
\end{equation}
we see that the influence of the substrate now enters via the torque term $\langle \boldsymbol{s}_0 \times \boldsymbol{S} \rangle$, which cannot be decoupled for the quantum spin without additional approximations, but must give origin to the double-frequency response.

\begin{figure}
\includegraphics[width=1.0\columnwidth]{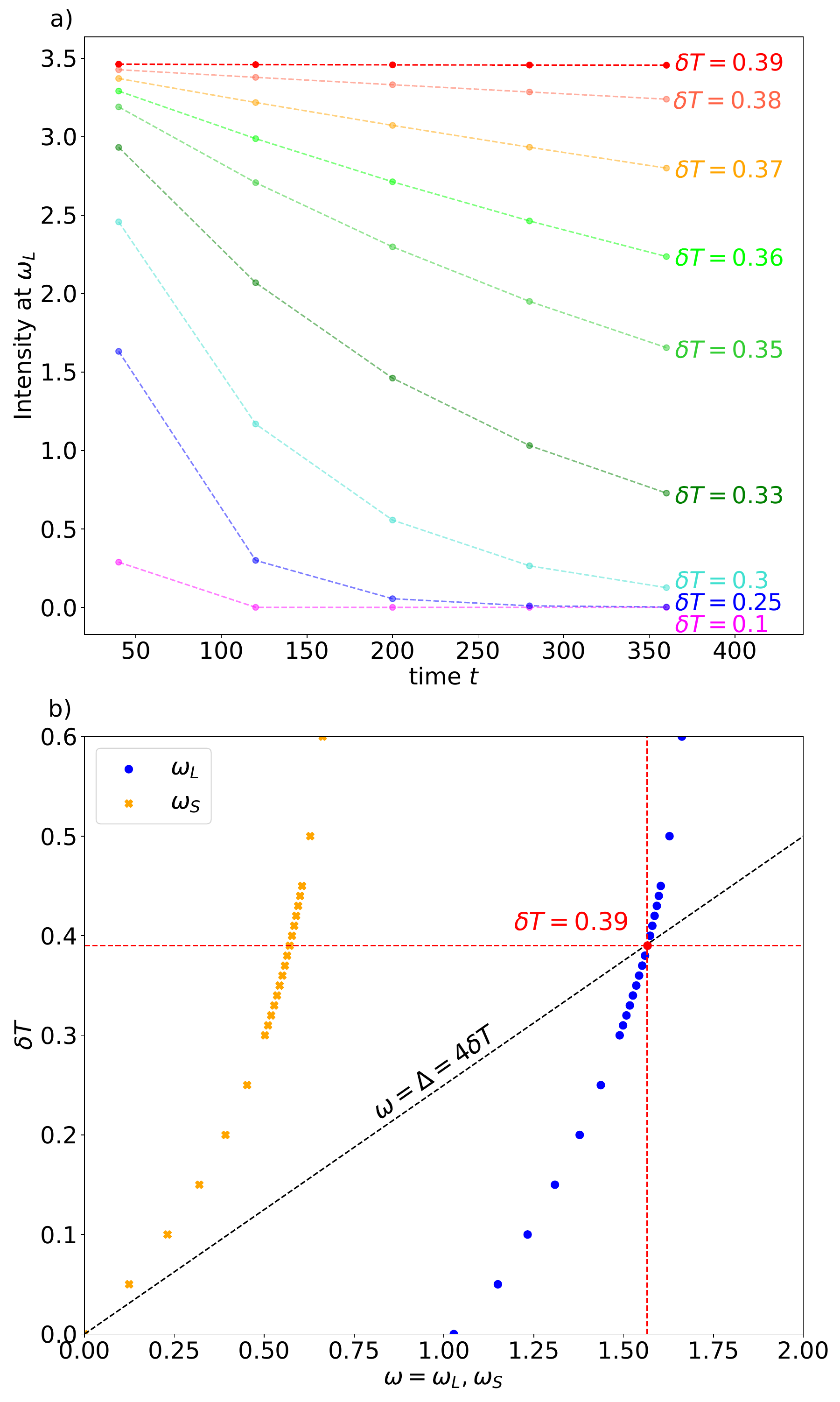}
\caption{(a) Intensity of the peak at the higher frequency $\omega_L$ at $B=1$ for different $\delta T$ and varying time windows (see Fig.~\ref{fig:Freq_Simp_delta}). (b) Dependence of two main frequencies $\omega_{L,S}$ on $\delta T$. The black dashed line shows $\omega=\Delta = 4 \big|\delta T\big|$.
}
\label{fig:Pre-relaxation}
\end{figure}

By varying $\delta T$, we observe that the higher frequency $\omega_L$ is continuously connected to the Larmor frequency $\omega_L\approx B$ at $\delta T=0$, see Fig.~\ref{fig:Pre-relaxation}(b). In Fig.~\ref{fig:Pre-relaxation}(a), we study the damping of the $\omega_L$ oscillation for various values of $\delta T$, and observe that within the accessible time window the damping ceases around $\delta T\approx 0.39$. If one employs the same criterion for relaxation as in the topologically trivial case $\omega_L = \Delta = 4\big|\delta T\big|$, one similarly finds $\delta T\approx0.39$ just by inserting the larger $\omega_L$ from the topological case, see Fig.~\ref{fig:Pre-relaxation}(b).

\begin{figure}
\includegraphics[width=1.0\columnwidth]{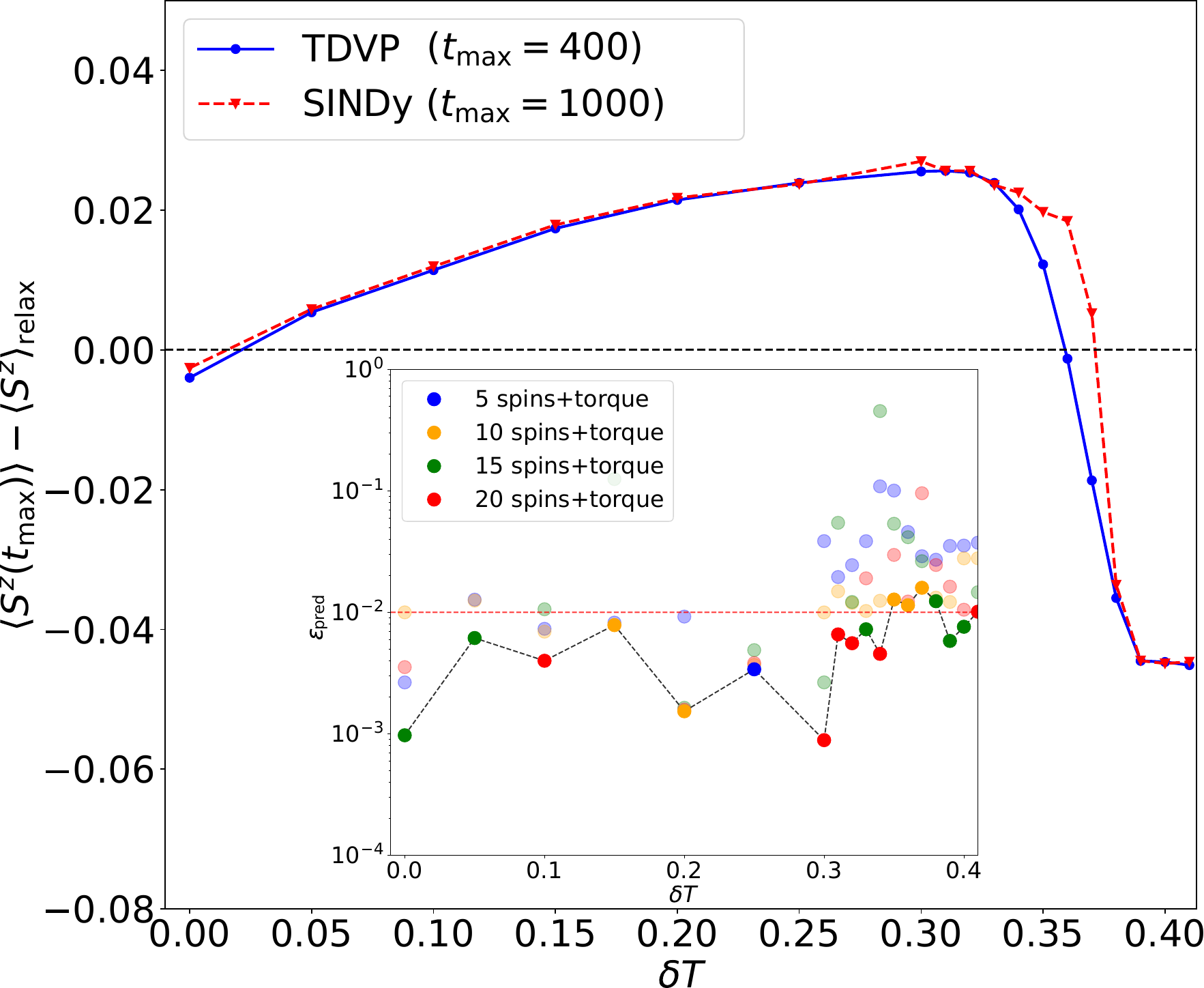}
\caption{Deviation of $\avg{S^z(t_{\text{max}})}$ from the expected relaxed value $\avg{S^z}_{\text{relax}}$ as a function of $\delta T$ at $B=1$. A transition from hyperpolarization to underpolarization occurs around $\delta T\sim 0.35$.
Inset: Average error Eq.~\eqref{eq: spin expectation error at tn} when taking a different number of $N_{\text{sub}}=5,10,15,20$ substrate spins into account within the SINDy approach. The black dashed line marks the minimal error.
} 
\label{fig:SINDy errorbar}
\end{figure}

In order to discern the pre-relaxation regime with hyperpolarization from the one with underpolarization, we analyze $\avg{S^z(t)}-\avg{S^z}_{\text{relax}}$ at the maximum time $t_\text{max}=400$ reached in the TDVP simulation. Figure \ref{fig:SINDy errorbar} shows this quantity as a function of $\delta T$ at $B=1$, indicating a transition around $\delta T\sim 0.35$.

One can obtain a better estimate for the boundary between the regimes of hyperpolarization and underpolarization by employing the SINDy method. We find that in the topologically non-trivial case, more observables are required than in the trivial case, and selecting these is not straightforward. To select them, we go back to the equation of motion for the impurity spin Eq.~\ref{eq: equation of motion of impurity spin}. We find that adding the torque term $\boldsymbol{T}(t) = \langle \boldsymbol{s}_0 \times \boldsymbol{S} \rangle$ as an additional observable to the list $\boldsymbol{A}(t)$ improves the quality of the SINDy prediction. Moreover, we find that it is beneficial to also include the first $N_{\text{sub}}$ substrate spins $\avg{\boldsymbol{s}_{i}}$. Hence, the matrix $\underline{\Theta}$ is now given by:
\begin{align}
 \underline{\Theta}(\underline{X}) &= 
  \begin{bmatrix}
    \mid &\mid &\mid& &\mid\\
    \langle \boldsymbol{S} \rangle & \langle \boldsymbol{T} \rangle&\langle \boldsymbol{s}_{0} \rangle&\dots&\langle \boldsymbol{s}_{N_{\text{sub}}-1} \rangle\\
    \mid & \mid & \mid& &\mid
  \end{bmatrix},\label{eq:function basis}\\
  \langle \boldsymbol{S} \rangle &= 
  \begin{bmatrix}
    \mid &\mid &\mid\\
    \langle \boldsymbol{S}^{x} \rangle & \langle  \boldsymbol{S}^{y} \rangle&\langle  \boldsymbol{S}^{z} \rangle\\
    \mid & \mid & \mid
  \end{bmatrix}, \\
  \langle \boldsymbol{T} \rangle &= 
  \begin{bmatrix}
    \mid &\mid &\mid\\
    \langle \boldsymbol{T}^{x} \rangle & \langle  \boldsymbol{T}^{y} \rangle&\langle  \boldsymbol{T}^{z} \rangle\\
    \mid & \mid & \mid
  \end{bmatrix},\\
    \langle \boldsymbol{s}_{j} \rangle &= 
  \begin{bmatrix}
    \mid &\mid &\mid\\
    \langle \boldsymbol{s}^{x}_{j} \rangle & \langle  \boldsymbol{s}^{y}_{j} \rangle&\langle  \boldsymbol{s}^{z}_{j} \rangle\\
    \mid & \mid & \mid
  \end{bmatrix}.
\end{align}

\begin{figure}
\includegraphics[width=1.0\columnwidth]{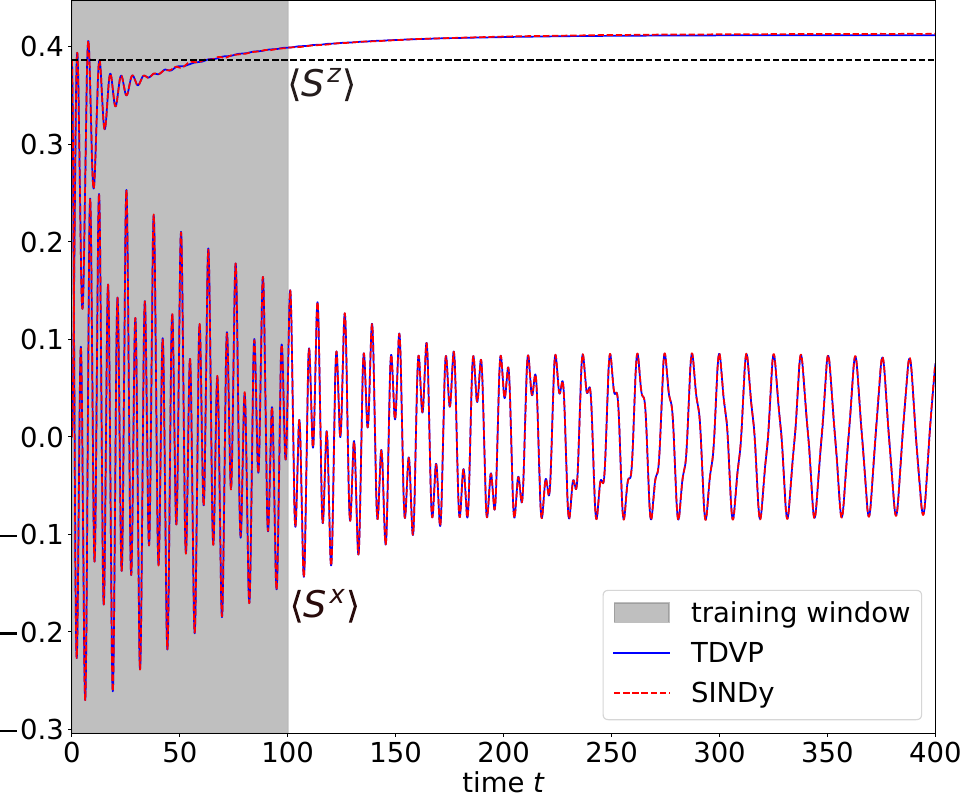}
\caption{Benchmark of the SINDy method for the topologically non-trivial case $\delta T=0.3$ at $B=1$. The impurity spin dynamics is learned from the TDVP results up to $t=100$ and is then used to predict the data up to $t=400$. We have included the expectation values of the impurity spin, the torque, and the first $N_{\text{sub}}=20$ substrate spins into the learning data.}
\label{fig:SINDy predictions for dT = 0.3}
\end{figure}

We find that using $t_\text{learn}=100$, i.e., 1/4 of the maximum TDVP simulation time, as the learning window allows us predict the remaining 3/4 of the data with good accuracy, which is exemplified in Fig.~\ref{fig:SINDy predictions for dT = 0.3}. In order to quantify how many substrate spins should be included in the SINDy approach, we introduce the relative prediction error at a given timestep as:
\begin{equation}
    \varepsilon_{\text{pred}}(t_n) = \Bigg|\frac{\langle \boldsymbol{S}_{\text{simul}}(t_n)\rangle-\langle\boldsymbol{S}_{\text{pred}}(t_n)\rangle}{\langle\boldsymbol{S}_{\text{simul}}(t_n)\rangle}\Bigg|,
\end{equation}
where the subscripts `simul' and `pred' refer to the TDVP data and the SINDy prediction, respectively. The total error is given by:
\begin{equation}
    \varepsilon_{\text{pred}} = \frac{1}{N} \sum_{n=0}^N \varepsilon_{\text{pred}}(t_{\text{learn}}+ndt).
\label{eq: spin expectation error at tn}
\end{equation}
In the inset to Fig.~\ref{fig:SINDy errorbar}, we plot $\epsilon_\text{pred}$ as a function of $\delta T$ at $B=1$. In most cases, using $N_{\text{sub}}=20$ spins yields the smallest error. From now on, the value of $N_\text{sub}\leq20$ is always chosen such that the error becomes minimal.

We now use the entire TDVP data up to $t_\text{max}=400$ to learn the dynamics and extract a prediction up to $t_\text{max}=1000$. We repeat the previous analysis to determine the the crossover between regimes of under- and hyperpolarization. The larger timescale accessible using the SINDy approach allows us to estimate the transition point more accurately, see Fig.~\ref{fig:SINDy errorbar}. The corresponding results are shown as the blue dashed line in Fig.~\ref{fig:regimes}.  We note that if one wanted to extend the timescale of the TDVP simulation to $t_\text{max}=1000$, one would have to employ a system size of $L\sim1000$, which would increase the numerical effort by at least a factor of $(2.5)^2\sim6$. This illustrates the potential of the SINDy method.

In the topologically nontrivial case, the counterintuitive regime of hyperpolarization cannot manifest for a classical spin. In the quantum case, the Kondo effect leads to a partial screening of the spin in the initial state, which hence allows for hyperpolarization. This can be compared to the phenomenon of ``quantum distillation'', where a low-entropy state with large double occupancy is dynamically created upon lifting a trapping potential for interacting fermions~\cite{Heidrich-Meisner2009}, which could be regarded as hyperpolarization of the double occupancy.

%%%%%%%%%%
\section{Discussion and Conclusion}
%%%%%%%%%%

We have studied the relaxation process of a quantum spin coupled to a fermionic substrate which we modelled as a one-dimensional SSH chain with a dimerization $\delta T$. The system was prepared in the ground state governed by a local magnetic field whose direction was switched at time $t=0$.

In the topologically trivial case, the quantum spin behaves similarly to a classical one; it relaxes if the Lamor frequency $\omega_L\approx B$ is larger than the single-particle gap $\Delta=4|\delta T|$ and does not relax otherwise.

In the topologically non-trivial case, we find three distinct regimes. For large $\delta T$, the spin does not relax; the corresponding region of the phase diagram is much larger than in the case of a classical spin. For small-to-intermediate $\delta T$, we find two distint pre-relaxation regimes: for small (intermediate) $\delta T$, the spin relaxes towards a value that is larger (smaller) than the expected ground-state value. We find that the transition between both regimes if fairly sharp. The Kondo effect leads to a polarization of the quantum spin below the maximally polarized state, opening the possibility of overshooting the relaxed value and approaching it from above. Our results can hence be viewed as a dynamical overcoming of the Kondo effect. It remains an open question if and how full relaxation occurs at very long timescales.

We find that the relaxation process is governed by two different frequencies. The larger frequency $\omega_L$ is adibatically connected to the Lamor frequency at $\delta T=0$ but becomes anomalously large with increasing $\delta T$. The oscillation with $\omega_L$ is damped for parameters $\Delta=4\delta T<\omega_L$, which lies in between the regimes of hyperpolarization and nonrelaxation. Nonrelaxation itself can be attributed to an anomalously high Lamor frequency. While the origin of this large Larmor frequency still needs to be clarified, it entails a well-defined energy scale, and some of the system behavior can understood by just comparing it to the band gap. We also note that an anomalously high Larmor frequency was observed before in the case of adiabatic dynamics for spins $S>1/2$~\cite{Stahl2017}.

Our results were obtained using charge-SU(2) symmetric tensor network algorithms, specifically the ground-state DMRG as well as the time-dependent variational principle. We subsequently employed the SINDy method to predict the impurity-spin dynamics from an initial learning window. This method can be combined with any propagation algorithm and is also applicable to general system-bath models. We find that the simple damping dynamics of the metallic case $\delta T=0$ is easily learnable; we can obtain an effective classical differential equation involving only $\avg{\mathbf{S}(t)}$, which can be regarded as a generalized Landau-Lifshitz-Gilbert equation. For a topologically nontrivial bath, no such simple equation can be found, and including system-bath correlations as well as bath-only expectation values becomes important. In this case, adding some domain knowledge of the physical problem is necessary in order to select the useful observables. One such guideline is to use the terms in the equation of motion for the system observables (in our case, the torque term). The SINDy approach allowed us to determine the crossover between the regimes of hyper- and underpolarization more accurately.

\subsection*{Acknowledgements}
Discussions with Michael Potthoff are gratefully acknowledged. This work was supported by the Deutsche Forschungsgemeinschaft through the grant KA 3360/4-1 (project number 508440990).

%%%%%%%%%
\begin{appendix}
%%%%%%%%%

\begin{figure}[b]
\includegraphics[width=\columnwidth]{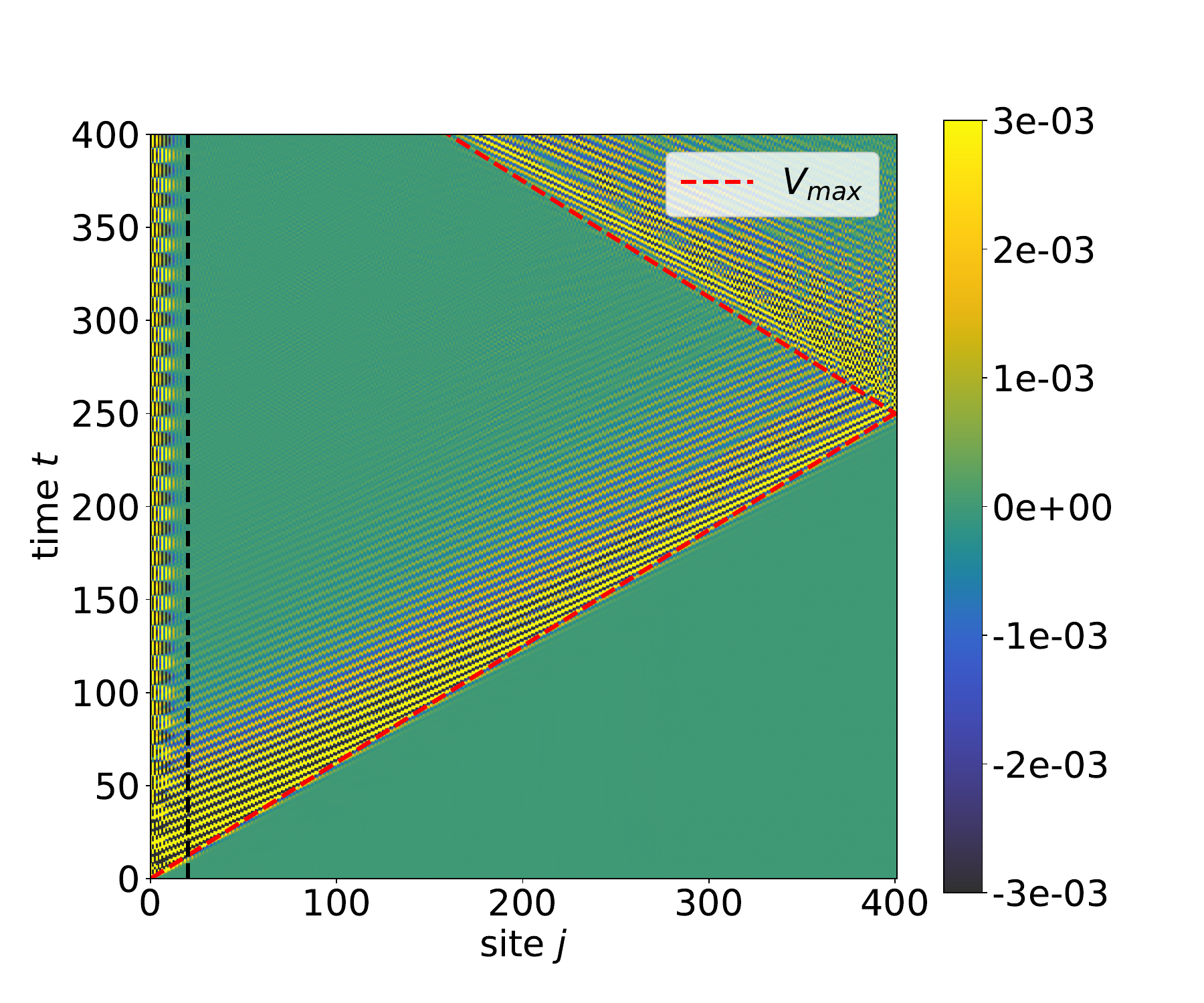}
\caption{Spin density $\avg{s^z_j(t)}$ in the substrate for $\delta T=0.3$ and $B=1$. The maximum wavefront speed calculated via Eq.~\eqref{eq:wavefront}  is shown as the red dashed line.}
\label{fig:fermi velocity}
\end{figure}

%%%%%%%%%%%%%%%%%%%%%%%%%%
\section{\label{app:proptime}Wavefront and maximal propagation time}
%%%%%%%%%%%%%%%%%%%%%%%%%%
In Fig.~\ref{fig:fermi velocity}, we show the spin density in the substrate $\avg{s^z_j(t)}$ as a function of the time and position. Excess energy that is put into the system by the field switching is transported away. The wavefront has a maximal velocity $v_{\text{max}}$, which can be estimated from
\begin{equation}
\begin{split}
v_{\text{max}} &= \max_{k} v(k) = \max_{k} \frac{\partial \epsilon(k)}{\partial k} \\
&=\max_{k} \frac{\partial}{\partial k} \sqrt{T_1^2+T_2^2+2T_1T_2\text{cos}(2k)}\\
&=\max_{k} \frac{2T_1T_2\text{sin}(2k)}{\sqrt{T_1^2+T_2^2+2T_1T_2\text{cos}(2k)}},
\label{eq:wavefront}
\end{split}
\end{equation}
where we used the dispersion relation of the free system $\epsilon(k)$, and $T_{1/2}=T\mp\delta T$.
Figure~\ref{fig:fermi velocity} shows that this describes the wavefront obtained in the numerics rather well. The wavefront is reflected at the opposite boundary of the chain.
To ensure that the impurity spin and substrate spins within $L_{\text{loc}}$ sites next to it are unperturbed by this reflection, we are at most allowed to propagate until the maximal time given by
\begin{equation}
    t_{\text{max}}(L,T_1,T_2) = \frac{2L-L_{\text{loc}}}{v_{\text{max}}(T_1,T_2)}.
\end{equation}

\section{\label{app:accuracy}Convergence test}

\begin{figure}[t]
\includegraphics[width=0.95\columnwidth]{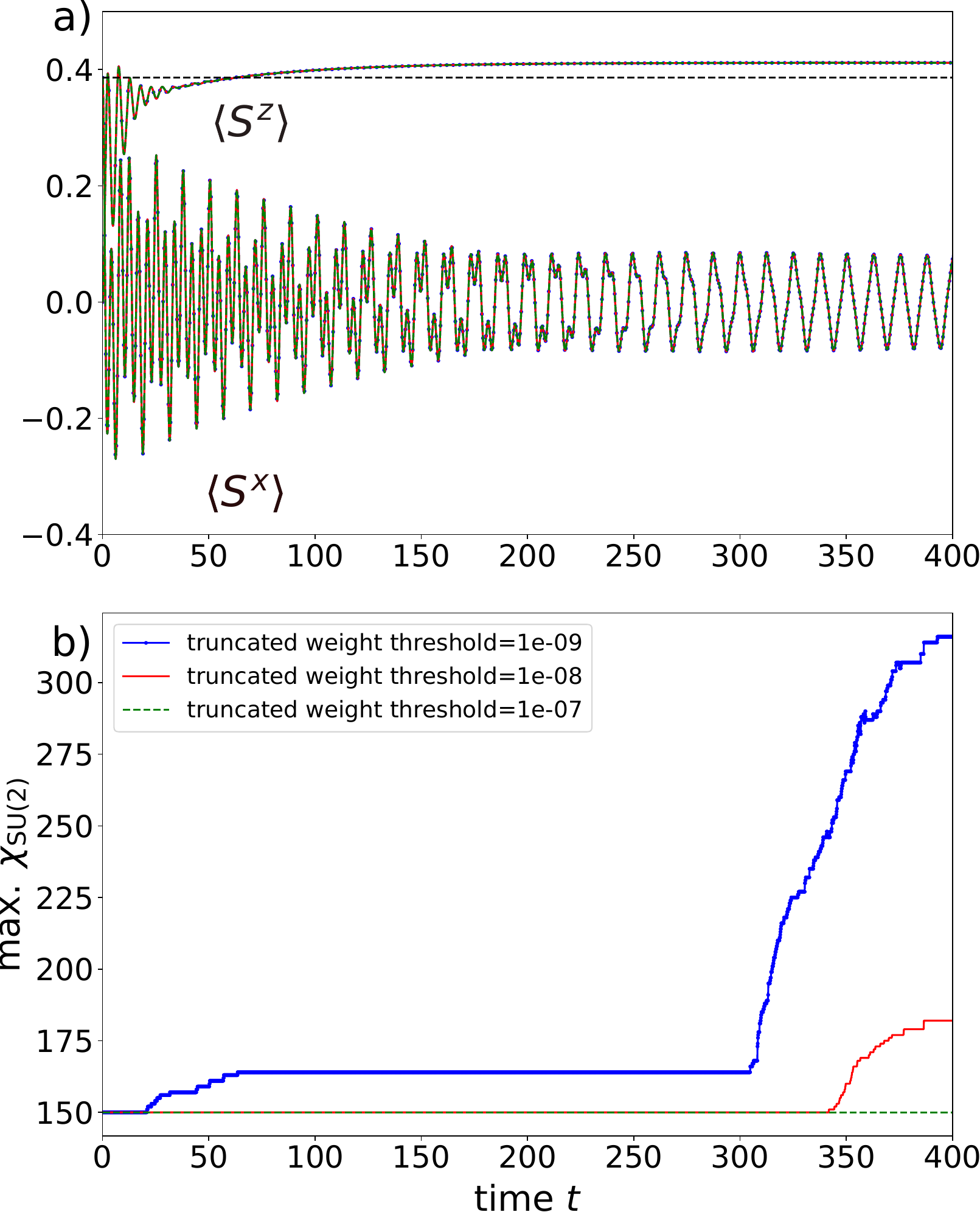}
\caption{TDVP time evolution for the topologically non-trivial case $\delta T=0.3$ at $B=1$ for three different values of the truncated weight. (a) Observables $\langle S^{x,z}\rangle$. (b) Maximum value of the effective bond dimensions $\chi_{\text{SU(2)}}$.
\label{fig:benchmark TDVP}}
\end{figure}

We briefly demonstrate that our TDVP data is converged w.r.t. the choice of the truncated weight. To this end, Fig.~\ref{fig:benchmark TDVP} shows the observables $\langle S^{x,z}\rangle$ as well as the bond dimension during a time evolution for some typical parameters.

\end{appendix}

\bibliography{ref}

\begin{thebibliography}{33}%
\makeatletter
\providecommand \@ifxundefined [1]{%
 \@ifx{#1\undefined}
}%
\providecommand \@ifnum [1]{%
 \ifnum #1\expandafter \@firstoftwo
 \else \expandafter \@secondoftwo
 \fi
}%
\providecommand \@ifx [1]{%
 \ifx #1\expandafter \@firstoftwo
 \else \expandafter \@secondoftwo
 \fi
}%
\providecommand \natexlab [1]{#1}%
\providecommand \enquote  [1]{``#1''}%
\providecommand \bibnamefont  [1]{#1}%
\providecommand \bibfnamefont [1]{#1}%
\providecommand \citenamefont [1]{#1}%
\providecommand \href@noop [0]{\@secondoftwo}%
\providecommand \href [0]{\begingroup \@sanitize@url \@href}%
\providecommand \@href[1]{\@@startlink{#1}\@@href}%
\providecommand \@@href[1]{\endgroup#1\@@endlink}%
\providecommand \@sanitize@url [0]{\catcode `\\12\catcode `\$12\catcode
  `\&12\catcode `\#12\catcode `\^12\catcode `\_12\catcode `\%12\relax}%
\providecommand \@@startlink[1]{}%
\providecommand \@@endlink[0]{}%
\providecommand \url  [0]{\begingroup\@sanitize@url \@url }%
\providecommand \@url [1]{\endgroup\@href {#1}{\urlprefix }}%
\providecommand \urlprefix  [0]{URL }%
\providecommand \Eprint [0]{\href }%
\providecommand \doibase [0]{https://doi.org/}%
\providecommand \selectlanguage [0]{\@gobble}%
\providecommand \bibinfo  [0]{\@secondoftwo}%
\providecommand \bibfield  [0]{\@secondoftwo}%
\providecommand \translation [1]{[#1]}%
\providecommand \BibitemOpen [0]{}%
\providecommand \bibitemStop [0]{}%
\providecommand \bibitemNoStop [0]{.\EOS\space}%
\providecommand \EOS [0]{\spacefactor3000\relax}%
\providecommand \BibitemShut  [1]{\csname bibitem#1\endcsname}%
\let\auto@bib@innerbib\@empty
%</preamble>
\bibitem [{\citenamefont {Kondo}(1969)}]{Kondo1969}%
  \BibitemOpen
  \bibfield  {author} {\bibinfo {author} {\bibfnamefont {J.}~\bibnamefont
  {Kondo}},\ }in\ \href@noop {} {\emph {\bibinfo {booktitle} {Solid State
  Physics}}},\ Vol.~\bibinfo {volume} {23},\ \bibinfo {editor} {edited by\
  \bibinfo {editor} {\bibfnamefont {F.}~\bibnamefont {Seitz}}, \bibinfo
  {editor} {\bibfnamefont {D.}~\bibnamefont {Turnbull}},\ and\ \bibinfo
  {editor} {\bibfnamefont {H.}~\bibnamefont {Ehrenreich}}}\ (\bibinfo
  {publisher} {Academic},\ \bibinfo {address} {New York},\ \bibinfo {year}
  {1969})\ p.\ \bibinfo {pages} {183}\BibitemShut {NoStop}%
\bibitem [{\citenamefont {Wilson}(1975)}]{Wilson1975}%
  \BibitemOpen
  \bibfield  {author} {\bibinfo {author} {\bibfnamefont {K.~G.}\ \bibnamefont
  {Wilson}},\ }\bibfield  {title} {\bibinfo {title} {The renormalization group:
  Critical phenomena and the kondo problem},\ }\href
  {https://doi.org/10.1103/RevModPhys.47.773} {\bibfield  {journal} {\bibinfo
  {journal} {Rev. Mod. Phys.}\ }\textbf {\bibinfo {volume} {47}},\ \bibinfo
  {pages} {773} (\bibinfo {year} {1975})}\BibitemShut {NoStop}%
\bibitem [{\citenamefont {Kohn}\ and\ \citenamefont
  {Santoro}(2022)}]{Kohn2021}%
  \BibitemOpen
  \bibfield  {author} {\bibinfo {author} {\bibfnamefont {L.}~\bibnamefont
  {Kohn}}\ and\ \bibinfo {author} {\bibfnamefont {G.~E.}\ \bibnamefont
  {Santoro}},\ }\bibfield  {title} {\bibinfo {title} {Quench dynamics of the
  anderson impurity model at finite temperature using matrix product states:
  entanglement and bath dynamics},\ }\href
  {https://doi.org/10.1088/1742-5468/ac729b} {\bibfield  {journal} {\bibinfo
  {journal} {Journal of Statistical Mechanics: Theory and Experiment}\ }\textbf
  {\bibinfo {volume} {2022}},\ \bibinfo {pages} {063102} (\bibinfo {year}
  {2022})}\BibitemShut {NoStop}%
\bibitem [{\citenamefont {Thoenniss}\ \emph {et~al.}(2023)\citenamefont
  {Thoenniss}, \citenamefont {Sonner}, \citenamefont {Lerose},\ and\
  \citenamefont {Abanin}}]{Thoenniss2023}%
  \BibitemOpen
  \bibfield  {author} {\bibinfo {author} {\bibfnamefont {J.}~\bibnamefont
  {Thoenniss}}, \bibinfo {author} {\bibfnamefont {M.}~\bibnamefont {Sonner}},
  \bibinfo {author} {\bibfnamefont {A.}~\bibnamefont {Lerose}},\ and\ \bibinfo
  {author} {\bibfnamefont {D.~A.}\ \bibnamefont {Abanin}},\ }\bibfield  {title}
  {\bibinfo {title} {Efficient method for quantum impurity problems out of
  equilibrium},\ }\href {https://doi.org/10.1103/PhysRevB.107.L201115}
  {\bibfield  {journal} {\bibinfo  {journal} {Phys. Rev. B}\ }\textbf {\bibinfo
  {volume} {107}},\ \bibinfo {pages} {L201115} (\bibinfo {year}
  {2023})}\BibitemShut {NoStop}%
\bibitem [{\citenamefont {Costi}(1997)}]{Costi1997}%
  \BibitemOpen
  \bibfield  {author} {\bibinfo {author} {\bibfnamefont {T.~A.}\ \bibnamefont
  {Costi}},\ }\bibfield  {title} {\bibinfo {title} {Renormalization-group
  approach to nonequilibrium green functionsin correlated impurity systems},\
  }\href {https://doi.org/10.1103/PhysRevB.55.3003} {\bibfield  {journal}
  {\bibinfo  {journal} {Phys. Rev. B}\ }\textbf {\bibinfo {volume} {55}},\
  \bibinfo {pages} {3003} (\bibinfo {year} {1997})}\BibitemShut {NoStop}%
\bibitem [{\citenamefont {Nordlander}\ \emph {et~al.}(1999)\citenamefont
  {Nordlander}, \citenamefont {Pustilnik}, \citenamefont {Meir}, \citenamefont
  {Wingreen},\ and\ \citenamefont {Langreth}}]{Nordlander1999}%
  \BibitemOpen
  \bibfield  {author} {\bibinfo {author} {\bibfnamefont {P.}~\bibnamefont
  {Nordlander}}, \bibinfo {author} {\bibfnamefont {M.}~\bibnamefont
  {Pustilnik}}, \bibinfo {author} {\bibfnamefont {Y.}~\bibnamefont {Meir}},
  \bibinfo {author} {\bibfnamefont {N.~S.}\ \bibnamefont {Wingreen}},\ and\
  \bibinfo {author} {\bibfnamefont {D.~C.}\ \bibnamefont {Langreth}},\
  }\bibfield  {title} {\bibinfo {title} {How long does it take for the kondo
  effect to develop?},\ }\href {https://doi.org/10.1103/PhysRevLett.83.808}
  {\bibfield  {journal} {\bibinfo  {journal} {Phys. Rev. Lett.}\ }\textbf
  {\bibinfo {volume} {83}},\ \bibinfo {pages} {808} (\bibinfo {year}
  {1999})}\BibitemShut {NoStop}%
\bibitem [{\citenamefont {Lechtenberg}\ and\ \citenamefont
  {Anders}(2014)}]{Lechtenberg2014}%
  \BibitemOpen
  \bibfield  {author} {\bibinfo {author} {\bibfnamefont {B.}~\bibnamefont
  {Lechtenberg}}\ and\ \bibinfo {author} {\bibfnamefont {F.~B.}\ \bibnamefont
  {Anders}},\ }\bibfield  {title} {\bibinfo {title} {Spatial and temporal
  propagation of kondo correlations},\ }\href
  {https://doi.org/10.1103/PhysRevB.90.045117} {\bibfield  {journal} {\bibinfo
  {journal} {Phys. Rev. B}\ }\textbf {\bibinfo {volume} {90}},\ \bibinfo
  {pages} {045117} (\bibinfo {year} {2014})}\BibitemShut {NoStop}%
\bibitem [{\citenamefont {Bouaziz}\ \emph {et~al.}(2019)\citenamefont
  {Bouaziz}, \citenamefont {Dias}, \citenamefont {Guimar\~aes},\ and\
  \citenamefont {Lounis}}]{Bouaziz2019}%
  \BibitemOpen
  \bibfield  {author} {\bibinfo {author} {\bibfnamefont {J.}~\bibnamefont
  {Bouaziz}}, \bibinfo {author} {\bibfnamefont {M.~d.~S.}\ \bibnamefont
  {Dias}}, \bibinfo {author} {\bibfnamefont {F.~S.~M.}\ \bibnamefont
  {Guimar\~aes}},\ and\ \bibinfo {author} {\bibfnamefont {S.}~\bibnamefont
  {Lounis}},\ }\bibfield  {title} {\bibinfo {title} {Spin dynamics of $3d$ and
  $4d$ impurities embedded in prototypical topological insulators},\ }\href
  {https://doi.org/10.1103/PhysRevMaterials.3.054201} {\bibfield  {journal}
  {\bibinfo  {journal} {Phys. Rev. Mater.}\ }\textbf {\bibinfo {volume} {3}},\
  \bibinfo {pages} {054201} (\bibinfo {year} {2019})}\BibitemShut {NoStop}%
\bibitem [{\citenamefont {Narayan}\ \emph {et~al.}(2015)\citenamefont
  {Narayan}, \citenamefont {Rungger},\ and\ \citenamefont
  {Sanvito}}]{Narayan2015}%
  \BibitemOpen
  \bibfield  {author} {\bibinfo {author} {\bibfnamefont {A.}~\bibnamefont
  {Narayan}}, \bibinfo {author} {\bibfnamefont {I.}~\bibnamefont {Rungger}},\
  and\ \bibinfo {author} {\bibfnamefont {S.}~\bibnamefont {Sanvito}},\
  }\bibfield  {title} {\bibinfo {title} {Single atom anisotropic
  magnetoresistance on a topological insulator surface},\ }\href
  {https://doi.org/10.1088/1367-2630/17/3/033021} {\bibfield  {journal}
  {\bibinfo  {journal} {New Journal of Physics}\ }\textbf {\bibinfo {volume}
  {17}},\ \bibinfo {pages} {033021} (\bibinfo {year} {2015})}\BibitemShut
  {NoStop}%
\bibitem [{\citenamefont {Chotorlishvili}\ \emph {et~al.}(2014)\citenamefont
  {Chotorlishvili}, \citenamefont {Ernst}, \citenamefont {Dugaev},
  \citenamefont {Komnik}, \citenamefont {Vergniory}, \citenamefont {Chulkov},\
  and\ \citenamefont {Berakdar}}]{Chotorlishvili2014}%
  \BibitemOpen
  \bibfield  {author} {\bibinfo {author} {\bibfnamefont {L.}~\bibnamefont
  {Chotorlishvili}}, \bibinfo {author} {\bibfnamefont {A.}~\bibnamefont
  {Ernst}}, \bibinfo {author} {\bibfnamefont {V.~K.}\ \bibnamefont {Dugaev}},
  \bibinfo {author} {\bibfnamefont {A.}~\bibnamefont {Komnik}}, \bibinfo
  {author} {\bibfnamefont {M.~G.}\ \bibnamefont {Vergniory}}, \bibinfo {author}
  {\bibfnamefont {E.~V.}\ \bibnamefont {Chulkov}},\ and\ \bibinfo {author}
  {\bibfnamefont {J.}~\bibnamefont {Berakdar}},\ }\bibfield  {title} {\bibinfo
  {title} {Magnetic fluctuations in topological insulators with ordered
  magnetic adatoms: Cr on bi${}_{2}$se${}_{3}$ from first principles},\ }\href
  {https://doi.org/10.1103/PhysRevB.89.075103} {\bibfield  {journal} {\bibinfo
  {journal} {Phys. Rev. B}\ }\textbf {\bibinfo {volume} {89}},\ \bibinfo
  {pages} {075103} (\bibinfo {year} {2014})}\BibitemShut {NoStop}%
\bibitem [{\citenamefont {Sarsen}\ and\ \citenamefont
  {Valagiannopoulos}(2019)}]{Sarsen2019}%
  \BibitemOpen
  \bibfield  {author} {\bibinfo {author} {\bibfnamefont {A.}~\bibnamefont
  {Sarsen}}\ and\ \bibinfo {author} {\bibfnamefont {C.}~\bibnamefont
  {Valagiannopoulos}},\ }\bibfield  {title} {\bibinfo {title} {Robust
  polarization twist by pairs of multilayers with tilted optical axes},\ }\href
  {https://doi.org/10.1103/PhysRevB.99.115304} {\bibfield  {journal} {\bibinfo
  {journal} {Phys. Rev. B}\ }\textbf {\bibinfo {volume} {99}},\ \bibinfo
  {pages} {115304} (\bibinfo {year} {2019})}\BibitemShut {NoStop}%
\bibitem [{\citenamefont {Hassani~Gangaraj}\ \emph {et~al.}(2020)\citenamefont
  {Hassani~Gangaraj}, \citenamefont {Valagiannopoulos},\ and\ \citenamefont
  {Monticone}}]{Gangaraj2020}%
  \BibitemOpen
  \bibfield  {author} {\bibinfo {author} {\bibfnamefont {S.~A.}\ \bibnamefont
  {Hassani~Gangaraj}}, \bibinfo {author} {\bibfnamefont {C.}~\bibnamefont
  {Valagiannopoulos}},\ and\ \bibinfo {author} {\bibfnamefont {F.}~\bibnamefont
  {Monticone}},\ }\bibfield  {title} {\bibinfo {title} {Topological scattering
  resonances at ultralow frequencies},\ }\href
  {https://doi.org/10.1103/PhysRevResearch.2.023180} {\bibfield  {journal}
  {\bibinfo  {journal} {Phys. Rev. Res.}\ }\textbf {\bibinfo {volume} {2}},\
  \bibinfo {pages} {023180} (\bibinfo {year} {2020})}\BibitemShut {NoStop}%
\bibitem [{\citenamefont {Su}\ \emph {et~al.}(1980)\citenamefont {Su},
  \citenamefont {Schrieffer},\ and\ \citenamefont {Heeger}}]{Su1980}%
  \BibitemOpen
  \bibfield  {author} {\bibinfo {author} {\bibfnamefont {W.~P.}\ \bibnamefont
  {Su}}, \bibinfo {author} {\bibfnamefont {J.~R.}\ \bibnamefont {Schrieffer}},\
  and\ \bibinfo {author} {\bibfnamefont {A.~J.}\ \bibnamefont {Heeger}},\
  }\bibfield  {title} {\bibinfo {title} {Soliton excitations in
  polyacetylene},\ }\href {https://doi.org/10.1103/PhysRevB.22.2099} {\bibfield
   {journal} {\bibinfo  {journal} {Phys. Rev. B}\ }\textbf {\bibinfo {volume}
  {22}},\ \bibinfo {pages} {2099} (\bibinfo {year} {1980})}\BibitemShut
  {NoStop}%
\bibitem [{\citenamefont {Heeger}\ \emph {et~al.}(1988)\citenamefont {Heeger},
  \citenamefont {Kivelson}, \citenamefont {Schrieffer},\ and\ \citenamefont
  {Su}}]{Heeger1988}%
  \BibitemOpen
  \bibfield  {author} {\bibinfo {author} {\bibfnamefont {A.~J.}\ \bibnamefont
  {Heeger}}, \bibinfo {author} {\bibfnamefont {S.}~\bibnamefont {Kivelson}},
  \bibinfo {author} {\bibfnamefont {J.~R.}\ \bibnamefont {Schrieffer}},\ and\
  \bibinfo {author} {\bibfnamefont {W.~P.}\ \bibnamefont {Su}},\ }\bibfield
  {title} {\bibinfo {title} {Solitons in conducting polymers},\ }\href
  {https://doi.org/10.1103/RevModPhys.60.781} {\bibfield  {journal} {\bibinfo
  {journal} {Rev. Mod. Phys.}\ }\textbf {\bibinfo {volume} {60}},\ \bibinfo
  {pages} {781} (\bibinfo {year} {1988})}\BibitemShut {NoStop}%
\bibitem [{\citenamefont {Elbracht}\ and\ \citenamefont
  {Potthoff}(2021)}]{Elbracht2021}%
  \BibitemOpen
  \bibfield  {author} {\bibinfo {author} {\bibfnamefont {M.}~\bibnamefont
  {Elbracht}}\ and\ \bibinfo {author} {\bibfnamefont {M.}~\bibnamefont
  {Potthoff}},\ }\bibfield  {title} {\bibinfo {title} {Long-time relaxation
  dynamics of a spin coupled to a chern insulator},\ }\href
  {https://doi.org/10.1103/physrevb.103.024301} {\bibfield  {journal} {\bibinfo
   {journal} {Physical Review B}\ }\textbf {\bibinfo {volume} {103}},\ \bibinfo
  {pages} {024301} (\bibinfo {year} {2021})}\BibitemShut {NoStop}%
\bibitem [{\citenamefont {Haegeman}\ \emph {et~al.}(2016)\citenamefont
  {Haegeman}, \citenamefont {Lubich}, \citenamefont {Oseledets}, \citenamefont
  {Vandereycken},\ and\ \citenamefont {Verstraete}}]{Haegeman2016}%
  \BibitemOpen
  \bibfield  {author} {\bibinfo {author} {\bibfnamefont {J.}~\bibnamefont
  {Haegeman}}, \bibinfo {author} {\bibfnamefont {C.}~\bibnamefont {Lubich}},
  \bibinfo {author} {\bibfnamefont {I.}~\bibnamefont {Oseledets}}, \bibinfo
  {author} {\bibfnamefont {B.}~\bibnamefont {Vandereycken}},\ and\ \bibinfo
  {author} {\bibfnamefont {F.}~\bibnamefont {Verstraete}},\ }\bibfield  {title}
  {\bibinfo {title} {Unifying time evolution and optimization with matrix
  product states},\ }\href {https://doi.org/10.1103/PhysRevB.94.165116}
  {\bibfield  {journal} {\bibinfo  {journal} {Phys. Rev. B}\ }\textbf {\bibinfo
  {volume} {94}},\ \bibinfo {pages} {165116} (\bibinfo {year}
  {2016})}\BibitemShut {NoStop}%
\bibitem [{\citenamefont {Brunton}\ \emph {et~al.}(2016)\citenamefont
  {Brunton}, \citenamefont {Proctor},\ and\ \citenamefont
  {Kutz}}]{Brunton2016}%
  \BibitemOpen
  \bibfield  {author} {\bibinfo {author} {\bibfnamefont {S.~L.}\ \bibnamefont
  {Brunton}}, \bibinfo {author} {\bibfnamefont {J.~L.}\ \bibnamefont
  {Proctor}},\ and\ \bibinfo {author} {\bibfnamefont {J.~N.}\ \bibnamefont
  {Kutz}},\ }\bibfield  {title} {\bibinfo {title} {Discovering governing
  equations from data by sparse identification of nonlinear dynamical
  systems},\ }\href {https://doi.org/10.1073/pnas.1517384113} {\bibfield
  {journal} {\bibinfo  {journal} {Proceedings of the National Academy of
  Sciences}\ }\textbf {\bibinfo {volume} {113}},\ \bibinfo {pages} {3932}
  (\bibinfo {year} {2016})}\BibitemShut {NoStop}%
\bibitem [{\citenamefont {Quade}\ \emph {et~al.}(2018)\citenamefont {Quade},
  \citenamefont {Abel}, \citenamefont {Nathan~Kutz},\ and\ \citenamefont
  {Brunton}}]{Quade2018}%
  \BibitemOpen
  \bibfield  {author} {\bibinfo {author} {\bibfnamefont {M.}~\bibnamefont
  {Quade}}, \bibinfo {author} {\bibfnamefont {M.}~\bibnamefont {Abel}},
  \bibinfo {author} {\bibfnamefont {J.}~\bibnamefont {Nathan~Kutz}},\ and\
  \bibinfo {author} {\bibfnamefont {S.~L.}\ \bibnamefont {Brunton}},\
  }\bibfield  {title} {\bibinfo {title} {Sparse identification of nonlinear
  dynamics for rapid model recovery},\ }\href
  {https://doi.org/10.1063/1.5027470} {\bibfield  {journal} {\bibinfo
  {journal} {Chaos: An Interdisciplinary Journal of Nonlinear Science}\
  }\textbf {\bibinfo {volume} {28}},\ \bibinfo {pages} {063116} (\bibinfo
  {year} {2018})}\BibitemShut {NoStop}%
\bibitem [{\citenamefont {Kaheman}\ \emph {et~al.}(2020)\citenamefont
  {Kaheman}, \citenamefont {Kutz},\ and\ \citenamefont
  {Brunton}}]{Kaheman2020}%
  \BibitemOpen
  \bibfield  {author} {\bibinfo {author} {\bibfnamefont {K.}~\bibnamefont
  {Kaheman}}, \bibinfo {author} {\bibfnamefont {J.~N.}\ \bibnamefont {Kutz}},\
  and\ \bibinfo {author} {\bibfnamefont {S.~L.}\ \bibnamefont {Brunton}},\
  }\bibfield  {title} {\bibinfo {title} {{SINDy-PI}: a robust algorithm for
  parallel implicit sparse identification of nonlinear dynamics},\ }\href
  {https://doi.org/10.1098/rspa.2020.0279} {\bibfield  {journal} {\bibinfo
  {journal} {Proceedings of the Royal Society A: Mathematical, Physical and
  Engineering Sciences}\ }\textbf {\bibinfo {volume} {476}},\ \bibinfo {pages}
  {20200279} (\bibinfo {year} {2020})}\BibitemShut {NoStop}%
\bibitem [{\citenamefont {Abdullah}\ and\ \citenamefont
  {Christofides}(2023)}]{Abdullah2023}%
  \BibitemOpen
  \bibfield  {author} {\bibinfo {author} {\bibfnamefont {F.}~\bibnamefont
  {Abdullah}}\ and\ \bibinfo {author} {\bibfnamefont {P.~D.}\ \bibnamefont
  {Christofides}},\ }\bibfield  {title} {\bibinfo {title} {Data-based modeling
  and control of nonlinear process systems using sparse identification: An
  overview of recent results},\ }\href
  {https://doi.org/https://doi.org/10.1016/j.compchemeng.2023.108247}
  {\bibfield  {journal} {\bibinfo  {journal} {Computers \& Chemical
  Engineering}\ }\textbf {\bibinfo {volume} {174}},\ \bibinfo {pages} {108247}
  (\bibinfo {year} {2023})}\BibitemShut {NoStop}%
\bibitem [{\citenamefont {Zhang}(1990)}]{Zhang1990}%
  \BibitemOpen
  \bibfield  {author} {\bibinfo {author} {\bibfnamefont {S.}~\bibnamefont
  {Zhang}},\ }\bibfield  {title} {\bibinfo {title} {Pseudospin symmetry and new
  collective modes of the {H}ubbard model},\ }\href
  {https://doi.org/10.1103/PhysRevLett.65.120} {\bibfield  {journal} {\bibinfo
  {journal} {Phys. Rev. Lett.}\ }\textbf {\bibinfo {volume} {65}},\ \bibinfo
  {pages} {120} (\bibinfo {year} {1990})}\BibitemShut {NoStop}%
\bibitem [{\citenamefont {Essler}\ \emph {et~al.}(2005)\citenamefont {Essler},
  \citenamefont {Frahm}, \citenamefont {G{\"o}hmann}, \citenamefont
  {Kl{\"u}mper},\ and\ \citenamefont {Korepin}}]{Essler2005}%
  \BibitemOpen
  \bibfield  {author} {\bibinfo {author} {\bibfnamefont {F.~H.}\ \bibnamefont
  {Essler}}, \bibinfo {author} {\bibfnamefont {H.}~\bibnamefont {Frahm}},
  \bibinfo {author} {\bibfnamefont {F.}~\bibnamefont {G{\"o}hmann}}, \bibinfo
  {author} {\bibfnamefont {A.}~\bibnamefont {Kl{\"u}mper}},\ and\ \bibinfo
  {author} {\bibfnamefont {V.~E.}\ \bibnamefont {Korepin}},\ }\href@noop {}
  {\emph {\bibinfo {title} {The one-dimensional {H}ubbard model}}}\ (\bibinfo
  {publisher} {Cambridge University Press},\ \bibinfo {year}
  {2005})\BibitemShut {NoStop}%
\bibitem [{\citenamefont {McCulloch}\ and\ \citenamefont
  {Gulácsi}(2002)}]{McCulloch2002}%
  \BibitemOpen
  \bibfield  {author} {\bibinfo {author} {\bibfnamefont {I.~P.}\ \bibnamefont
  {McCulloch}}\ and\ \bibinfo {author} {\bibfnamefont {M.}~\bibnamefont
  {Gulácsi}},\ }\bibfield  {title} {\bibinfo {title} {The non-abelian density
  matrix renormalization group algorithm},\ }\href
  {https://doi.org/10.1209/epl/i2002-00393-0} {\bibfield  {journal} {\bibinfo
  {journal} {Europhysics Letters}\ }\textbf {\bibinfo {volume} {57}},\ \bibinfo
  {pages} {852} (\bibinfo {year} {2002})}\BibitemShut {NoStop}%
\bibitem [{\citenamefont {Hubig}\ \emph {et~al.}(2015)\citenamefont {Hubig},
  \citenamefont {McCulloch}, \citenamefont {Schollw\"ock},\ and\ \citenamefont
  {Wolf}}]{Hubig2015}%
  \BibitemOpen
  \bibfield  {author} {\bibinfo {author} {\bibfnamefont {C.}~\bibnamefont
  {Hubig}}, \bibinfo {author} {\bibfnamefont {I.~P.}\ \bibnamefont
  {McCulloch}}, \bibinfo {author} {\bibfnamefont {U.}~\bibnamefont
  {Schollw\"ock}},\ and\ \bibinfo {author} {\bibfnamefont {F.~A.}\ \bibnamefont
  {Wolf}},\ }\bibfield  {title} {\bibinfo {title} {Strictly single-site dmrg
  algorithm with subspace expansion},\ }\href
  {https://doi.org/10.1103/PhysRevB.91.155115} {\bibfield  {journal} {\bibinfo
  {journal} {Phys. Rev. B}\ }\textbf {\bibinfo {volume} {91}},\ \bibinfo
  {pages} {155115} (\bibinfo {year} {2015})}\BibitemShut {NoStop}%
\bibitem [{\citenamefont {Breuer}\ and\ \citenamefont
  {Petruccione}(2002)}]{Breuer2002}%
  \BibitemOpen
  \bibfield  {author} {\bibinfo {author} {\bibfnamefont {H.-P.}\ \bibnamefont
  {Breuer}}\ and\ \bibinfo {author} {\bibfnamefont {F.}~\bibnamefont
  {Petruccione}},\ }\href@noop {} {\emph {\bibinfo {title} {The theory of open
  quantum systems}}}\ (\bibinfo  {publisher} {OUP Oxford},\ \bibinfo {year}
  {2002})\BibitemShut {NoStop}%
\bibitem [{\citenamefont {Fransson}(2008)}]{Fransson2008}%
  \BibitemOpen
  \bibfield  {author} {\bibinfo {author} {\bibfnamefont {J.}~\bibnamefont
  {Fransson}},\ }\bibfield  {title} {\bibinfo {title} {Detection of spin
  reversal and nutations through current measurements},\ }\href
  {https://doi.org/10.1088/0957-4484/19/28/285714} {\bibfield  {journal}
  {\bibinfo  {journal} {Nanotechnology}\ }\textbf {\bibinfo {volume} {19}},\
  \bibinfo {pages} {285714} (\bibinfo {year} {2008})}\BibitemShut {NoStop}%
\bibitem [{\citenamefont {Bhattacharjee}\ \emph {et~al.}(2012)\citenamefont
  {Bhattacharjee}, \citenamefont {Nordstr\"om},\ and\ \citenamefont
  {Fransson}}]{Bhattacharjee2012}%
  \BibitemOpen
  \bibfield  {author} {\bibinfo {author} {\bibfnamefont {S.}~\bibnamefont
  {Bhattacharjee}}, \bibinfo {author} {\bibfnamefont {L.}~\bibnamefont
  {Nordstr\"om}},\ and\ \bibinfo {author} {\bibfnamefont {J.}~\bibnamefont
  {Fransson}},\ }\bibfield  {title} {\bibinfo {title} {Atomistic spin dynamic
  method with both damping and moment of inertia effects included from first
  principles},\ }\href {https://doi.org/10.1103/PhysRevLett.108.057204}
  {\bibfield  {journal} {\bibinfo  {journal} {Phys. Rev. Lett.}\ }\textbf
  {\bibinfo {volume} {108}},\ \bibinfo {pages} {057204} (\bibinfo {year}
  {2012})}\BibitemShut {NoStop}%
\bibitem [{\citenamefont {Sayad}\ and\ \citenamefont
  {Potthoff}(2015)}]{Sayad2015}%
  \BibitemOpen
  \bibfield  {author} {\bibinfo {author} {\bibfnamefont {M.}~\bibnamefont
  {Sayad}}\ and\ \bibinfo {author} {\bibfnamefont {M.}~\bibnamefont
  {Potthoff}},\ }\bibfield  {title} {\bibinfo {title} {Spin dynamics and
  relaxation in the classical-spin kondo-impurity model beyond the
  landau–lifschitz–gilbert equation},\ }\href
  {https://doi.org/10.1088/1367-2630/17/11/113058} {\bibfield  {journal}
  {\bibinfo  {journal} {New Journal of Physics}\ }\textbf {\bibinfo {volume}
  {17}},\ \bibinfo {pages} {113058} (\bibinfo {year} {2015})}\BibitemShut
  {NoStop}%
\bibitem [{\citenamefont {Vedmedenko}\ and\ \citenamefont
  {Potthoff}(2018)}]{Vedmedenko2018}%
  \BibitemOpen
  \bibfield  {author} {\bibinfo {author} {\bibfnamefont {E.}~\bibnamefont
  {Vedmedenko}}\ and\ \bibinfo {author} {\bibfnamefont {M.}~\bibnamefont
  {Potthoff}},\ }\bibinfo {title} {Fluctuations and dynamics of magnetic
  nanoparticles},\ in\ \href {https://doi.org/10.1007/978-3-319-99558-8_13}
  {\emph {\bibinfo {booktitle} {Atomic- and Nanoscale Magnetism}}},\ \bibinfo
  {editor} {edited by\ \bibinfo {editor} {\bibfnamefont {R.}~\bibnamefont
  {Wiesendanger}}}\ (\bibinfo  {publisher} {Springer International
  Publishing},\ \bibinfo {address} {Cham},\ \bibinfo {year} {2018})\ pp.\
  \bibinfo {pages} {267--284}\BibitemShut {NoStop}%
\bibitem [{\citenamefont {Elbracht}\ and\ \citenamefont
  {Potthoff}(2024)}]{Elbracht2024}%
  \BibitemOpen
  \bibfield  {author} {\bibinfo {author} {\bibfnamefont {M.}~\bibnamefont
  {Elbracht}}\ and\ \bibinfo {author} {\bibfnamefont {M.}~\bibnamefont
  {Potthoff}},\ }\bibfield  {title} {\bibinfo {title} {Pre-relaxation in
  quantum, classical, and quantum-classical two-impurity models},\ }\href
  {https://arxiv.org/abs/2404.18566} {\bibfield  {journal} {\bibinfo  {journal}
  {arXiv preprint}\ } (\bibinfo {year} {2024})}\BibitemShut {NoStop}%
\bibitem [{\citenamefont {Sayad}\ \emph {et~al.}(2016)\citenamefont {Sayad},
  \citenamefont {Rausch},\ and\ \citenamefont {Potthoff}}]{Sayad2016}%
  \BibitemOpen
  \bibfield  {author} {\bibinfo {author} {\bibfnamefont {M.}~\bibnamefont
  {Sayad}}, \bibinfo {author} {\bibfnamefont {R.}~\bibnamefont {Rausch}},\ and\
  \bibinfo {author} {\bibfnamefont {M.}~\bibnamefont {Potthoff}},\ }\bibfield
  {title} {\bibinfo {title} {Inertia effects in the real-time dynamics of a
  quantum spin coupled to a fermi sea},\ }\href
  {https://doi.org/10.1209/0295-5075/116/17001} {\bibfield  {journal} {\bibinfo
   {journal} {Europhysics Letters}\ }\textbf {\bibinfo {volume} {116}},\
  \bibinfo {pages} {17001} (\bibinfo {year} {2016})}\BibitemShut {NoStop}%
\bibitem [{\citenamefont {Heidrich-Meisner}\ \emph {et~al.}(2009)\citenamefont
  {Heidrich-Meisner}, \citenamefont {Manmana}, \citenamefont {Rigol},
  \citenamefont {Muramatsu}, \citenamefont {Feiguin},\ and\ \citenamefont
  {Dagotto}}]{Heidrich-Meisner2009}%
  \BibitemOpen
  \bibfield  {author} {\bibinfo {author} {\bibfnamefont {F.}~\bibnamefont
  {Heidrich-Meisner}}, \bibinfo {author} {\bibfnamefont {S.~R.}\ \bibnamefont
  {Manmana}}, \bibinfo {author} {\bibfnamefont {M.}~\bibnamefont {Rigol}},
  \bibinfo {author} {\bibfnamefont {A.}~\bibnamefont {Muramatsu}}, \bibinfo
  {author} {\bibfnamefont {A.~E.}\ \bibnamefont {Feiguin}},\ and\ \bibinfo
  {author} {\bibfnamefont {E.}~\bibnamefont {Dagotto}},\ }\bibfield  {title}
  {\bibinfo {title} {Quantum distillation: Dynamical generation of low-entropy
  states of strongly correlated fermions in an optical lattice},\ }\href
  {https://doi.org/10.1103/PhysRevA.80.041603} {\bibfield  {journal} {\bibinfo
  {journal} {Phys. Rev. A}\ }\textbf {\bibinfo {volume} {80}},\ \bibinfo
  {pages} {041603} (\bibinfo {year} {2009})}\BibitemShut {NoStop}%
\bibitem [{\citenamefont {Stahl}\ and\ \citenamefont
  {Potthoff}(2017)}]{Stahl2017}%
  \BibitemOpen
  \bibfield  {author} {\bibinfo {author} {\bibfnamefont {C.}~\bibnamefont
  {Stahl}}\ and\ \bibinfo {author} {\bibfnamefont {M.}~\bibnamefont
  {Potthoff}},\ }\bibfield  {title} {\bibinfo {title} {Anomalous spin
  precession under a geometrical torque},\ }\href
  {https://doi.org/10.1103/PhysRevLett.119.227203} {\bibfield  {journal}
  {\bibinfo  {journal} {Phys. Rev. Lett.}\ }\textbf {\bibinfo {volume} {119}},\
  \bibinfo {pages} {227203} (\bibinfo {year} {2017})}\BibitemShut {NoStop}%
\end{thebibliography}%

\end{document}